\documentclass[iop,revtex4]{emulateapj}

\shorttitle{THE QUEST FOR HIGH-Z DUSTY STARFORMING GALAXIES}
\shortauthors{C. MANCUSO ET AL.}
\slugcomment{ACCEPTED BY ApJ}

\begin{document}

\title{The Quest for Dusty Star-forming Galaxies at High Redshift $z\ga 4$}
\author{C. Mancuso\altaffilmark{1,2,3}, A. Lapi\altaffilmark{1,2,3}, J. Shi\altaffilmark{1,4}, J. Gonzalez-Nuevo\altaffilmark{5}, R. Aversa\altaffilmark{1,3}, L. Danese\altaffilmark{1,2,3}}
\altaffiltext{1}{SISSA, Via Bonomea 265, 34136 Trieste, Italy}
\altaffiltext{2}{INAF-Osservatorio Astronomico di Trieste, via Tiepolo 11, 34131 Trieste, Italy} \altaffiltext{3}{INFN-Sezione di
Trieste, via Valerio 2, 34127 Trieste, Italy}\altaffiltext{4}{Key Lab. for
Research in Galaxies and Cosmology, Dept. of Astronomy, Univ. of Science and
Technology of China, Hefei, 230026 Anhui, China}\altaffiltext{5}{Departamento de F\'isica, Universidad de Oviedo, C. Calvo Sotelo s/n, 33007 Oviedo, Spain}

\begin{abstract}
We exploit the continuity equation approach and the `main sequence'
star-formation timescales to show that the observed high abundance of galaxies with stellar masses $\ga$ a few $10^{10}\, M_{\odot}$ at redshift
$z\ga 4$ implies the existence of a galaxy population featuring large star formation rates (SFRs) $\psi\ga 10^2\, M_\odot$ yr$^{-1}$ in heavily dust-obscured conditions. These galaxies constitute the high-redshift counterparts of the dusty star-forming population already surveyed for $z\la 3$ in the far-IR band by the \textsl{Herschel} space observatory. We work out specific predictions for the evolution of the corresponding stellar mass and SFR functions out to $z\sim 10$, elucidating that the number density at $z\la 8$ for SFRs $\psi\ga 30\, M_\odot$ yr$^{-1}$ cannot be estimated relying on the UV luminosity function alone, even when standard corrections for dust extinction based on the UV slope are applied. We compute the number counts and redshift distributions (including galaxy-scale gravitational lensing) of this galaxy population, and show that current data from \textsl{AzTEC}-\textsl{LABOCA}, \textsl{SCUBA-2} and \textsl{ALMA}-\textsl{SPT} surveys are already digging into it. We substantiate how an observational strategy based on a color preselection in the far-IR or (sub-)mm band with \textsl{Herschel} and \textsl{SCUBA-2}, supplemented by photometric data via on-source observations with \textsl{ALMA}, can allow to reconstruct the bright end of the SFR functions out to $z\la 8$. In parallel, such a challenging task can be managed by exploiting current UV surveys in combination with (sub-)mm observations by \textsl{ALMA} and \textsl{NIKA2} and/or radio observations by \textsl{SKA} and its precursors.
\end{abstract}


\setcounter{footnote}{0}

\section{Introduction}\label{sec|intro}

The star formation in galaxies can be inferred by lines like Ly${\alpha}$ and H${\alpha}$, and by continuum emission in the ultraviolet (UV), infrared
(IR), radio and X-ray bands (see Kennicutt \& Evans 2012 for a review). In
the local Universe a significant fraction of the star formation in galaxies
occurs in dust-enshrouded environments (e.g., Carilli et al. 2013; Madau \&
Dickinson 2014), with a clear tendency for dust extinction to become more
severe as the star formation rate (SFR) increases. Dust causes the UV
emission from young massive stars, which traces the SFR, to be absorbed and
reradiated in the far-IR band; thus a combined measurement of UV and far-IR
luminosities would constitute a sound probe of the SFR.

Even at high redshift dusty star-forming galaxies are quite common, as shown
by the large surveys obtained by ground- and space-based instruments in the
recent years (for a review, see Casey et al. 2014). The tendency for dust
obscuration to increase with SFR is also confirmed by the increase of the
UV-continuum slope $\beta_{\rm UV}$ with raising luminosity in UV-selected galaxies up to $z\sim 8$ (see Bouwens et al. 2014, and references in their
Fig.~1; also Reddy et al. 2012; Coppin et al. 2015).

The correlation of the UV slope $\beta_{\rm UV}$ with the ratio of the IR to
UV luminosity (dubbed IRX ratio) in star-forming galaxies has been commonly
exploited in order to estimate their dust absorption (e.g., Meurer et al.
1999). As a matter of fact, far-IR observations of UV-selected galaxies
confirmed that the estimates of dust attenuation based on the $\beta_{\rm
UV}$-IRX correlation are reliable for objects with SFR $\psi\la 30\,
M_{\odot}$ yr$^{-1}$ (e.g., Lee et al. 2012; Reddy et al. 2012, 2015; Coppin
et al. 2015). On the other hand, the scatter of the $\beta_{\rm UV}$-IRX
relation largely widens with increasing $\beta_{\rm UV}$ and IRX (i.e., with
increasing SFR on the average), making the dust correction quite uncertain
for SFR $\psi\ga 30\, M_{\odot}$ yr$^{-1}$ (e.g., Chapman et al. 2000, Goldader et al. 2002; for a recent review, see Conroy 2013).

The relevance of dust absorption is evident from the shape and redshift
evolution of the luminosity function at the bright end (e.g., Mao et al.
2007; Bouwens et al. 2009; Cai et al. 2014; Bowler et al. 2015). More
precisely, the uncertainty in the dust absorption strongly affects the
estimate of the SFR function at the bright end, as inferred from UV surveys.
At redshift up to $z\sim 3$ the effect has been statistically quantified by
Aversa et al. (2015), by comparing the SFR function as inferred by the UV
luminosity function (corrected for the dust absorption basing on the UV
slope) with that inferred by the far-IR surveys obtained with the SPIRE
instrument on board of \textsl{Herschel} (see Lapi et al. 2011; Gruppioni et al. 2013, 2015; Magnelli et al. 2013). These authors have shown that UV surveys
start to undersample galaxies endowed with SFR $\psi\ga 30\, M_{\odot}$
yr$^{-1}$, even when corrected by dust attenuation on the basis of the UV
slope-IRX correlation. They have also highlighted that the galaxy stellar
mass function at $z\la 3$ can be recovered from the intrinsic SFR function.
At higher redshifts $z\ga 3$ the direct comparison is hampered by the fact
that, while the UV luminosity function is soundly determined up to redshift
$z\sim 8$ (e.g., Bouwens et al. 2015; Finkelstein et al. 2015a; Bowler et al. 2015), the far-IR luminosity function is not yet available due to the
sensitivity limits of current instruments.

To circumvent the problem, Aversa et al. (2015) have argued that the
continuity equation applied to SFR and stellar mass can provide an important
clue to the distribution of the intrinsic SFR even at $z\ga 4$. In this vein, it is worth noticing that the spectral energy distribution (SED) for large
samples of high redshift galaxies has been recently determined from the UV
to near-/mid-IR, allowing for sound estimate of the photometric redshift,
stellar mass, dust extinction, SFR and age of stellar populations (e.g.,
Duncan et al. 2014; Speagle et al. 2014; Salmon et al. 2015; Grazian et al.
2015; Caputi et al. 2015; Stefanon et al. 2015), with due caveats related to the degeneracy among these parameters (e.g., Conroy 2013).

Estimate of the galaxy stellar mass function at substantial redshift has been obtained by combining the observed mass-to-UV light ratio and the UV
luminosity function (e.g., Stark et al. 2009; Gonzalez et al. 2011; Lee et
al. 2012; Song et al. 2015). However, both these key ingredients are expected to be affected by dust extinction at high UV luminosity. Moreover, the
correlation $M_{\star}-M_{\rm UV}$ is largely scattered. Note that such a
relation is also relevant for the definition of the so called `main
sequence', once the UV luminosity is translated into SFR, provided that the
dust effects are properly taken into account.

Bypassing the UV selection, deep optical/near-IR/mid-IR imaging provided by the \textsl{HST}, \textsl{Spitzer}, and the \textsl{VLT} on CANDELS-UDS, GOODS-South, and HUDF fields have recently been exploited  in order to determine the galaxy stellar mass function at redshift $z\ga 3$, with the stellar mass derived from the SED fitting technique including nebular emission (Duncan et al. 2014; Grazian et al. 2015; also Caputi et al 2015). The stellar mass function has been computed by weighting the galaxies with the $1/V_{\rm max}$ Schmidt's estimator (Schmidt 1968). The outcome agrees with that derived only for UV-selected galaxies when large intrinsic scatter $\ga 0.5$ dex in the $M_{\star}-M_{\rm UV}$ relation is assumed (see Fig.~9 in Duncan et al. 2014). Such a wide scatter suggests that a fraction of the low luminosity UV-selected galaxies are already massive, and that either they are already quiescent or they form most of their stars within a dusty interstellar medium (ISM; see also Grazian et al. 2015). As a matter of fact, Song et al. (2015) notice the increase of massive but faint UV galaxies at lower redshift, suggesting that the role of the dust is increasingly relevant with cosmic time. Also Bowler et al. (2015) point out that the bright end of the UV luminosity function appears to steepen from $z\sim 7$ to $5$, possibly suggesting the onset of dust obscuration. Additional evidence for the presence of dust at quite high redshift is confirmed by observations of quasars (e.g., Bolton et al. 2011), direct detection from \textsl{ALMA} (e.g., Weiss et al. 2013; Swinbank et al. 2014; da Cunha et al. 2015) and indirectly from the nature of high-$z$ gamma ray bursts (e.g., Schady et al. 2014).

In this paper we aim at deriving a determination of the intrinsic SFR and
stellar mass functions, unbiased with respect to dust obscuration; these are
indeed crucial ingredients for our physical understanding of galaxy formation and evolution. For example, these functions can be used to obtain
\emph{intrinsic} relationships of the SFR/stellar mass vs. the dark matter halo mass via the abundance matching technique (e.g., Vale \& Ostriker 2004, Shankar et al. 2006, Moster et al. 2013; Behroozi et al. 2013). We shall see that the exploitation of such intrinsic relationships, as opposed to those derived after dust corrections based on the UV slope, leads to naturally solve a couple of critical issues in galaxy formation and evolution, pointed out by Steinhardt et al. (2015) and by Finkelstein et al. (2015b):
the former authors claim massive high-redshift galaxies to have formed impossibly early according to standard models of galaxy assembly;
the latter authors point out an unexpected increase of the stellar to baryon fraction in bright galaxies at high redshift.

The plan of the paper is the following: in Sect.~\ref{sec|SFRfunc} we develop a new method to obtain an analytic rendition of the intrinsic SFR function at any redshift $z\sim 0-10$ from the most recent UV and far-IR data. In
Sect.~\ref{sec|counts} we validate our intrinsic SFR function by comparison
with the observed (sub-)mm counts, redshift distributions, and cosmic infrared background. In Sect.~\ref{sec|cont} we exploit the
continuity equation approach to further validate our intrinsic SFR function
by comparison with the observed stellar mass function at high redshift $z\ga
4$. In Sect.~\ref{sec|abmatch} we use the abundance matching technique to
derive relationships between SFR and stellar mass vs. the halo mass, and
discuss their consequences for galaxy formation scenarios. In
Sect.~\ref{sec|hunting} we design specific observational strategies to hunt
for high-$z$ dusty galaxies that we predict to populate the bright end of the intrinsic SFR function for $z\ga 4$, by exploiting far-IR/(sub-)mm (Sect.~\ref{sec|FIRnature}) and/or UV surveys (Sect.~\ref{sec|UVnature}). Finally, in Sect.~\ref{sec|summary} we summarize our results.

Throughout this work we adopt the standard flat concordance cosmology (Planck Collaboration XIII 2015) with round parameter values: matter density $\Omega_M = 0.32$, baryon density $\Omega_b = 0.05$, Hubble constant $H_0 = 100\, h$ km s$^{−1}$ Mpc$^{−1}$ with $h = 0.67$, and mass variance $\sigma_8 = 0.83$ on a scale of $8\, h^{-1}$ Mpc. Stellar masses and luminosities (or SFRs) of galaxies are evaluated assuming the Chabrier's (2003) initial mass function (IMF).

\section{Reconstructing the intrinsic SFR function}\label{sec|SFRfunc}

From an observational point of view, the intrinsic SFR function $N(\log
\psi,z)$, namely the number of galaxies per logarithmic bin of SFR $[\log
\psi,\log\psi+{\rm d}\log\psi]$ at given redshift $z$, is mainly determined
from pure UV or pure far-IR selected samples; in both cases
\emph{corrections} come into play and must be taken into proper account to
infer the intrinsic SFR function. When basing solely on IR measurements, the
main issue concerns the contribution to the global IR luminosity coming from
diffuse dust (cirrus), that reprocesses the light from less massive, older
stars; in fact, the SFR is better traced by the dust emission from molecular
clouds, that instead reprocesses the UV light from young massive stars. Not
correcting the global luminosity for diffuse (cirrus) emission would cause the SFR to be appreciably overestimated; however, this is not an easy task since diffuse emission depends on several aspects like stellar mass, galaxy age, chemical composition, dust amount and related spatial distribution (see Silva et al. 1998). On the other hand, several studies in the local Universe (e.g., Hao et al. 2011; Clemens et al. 2013; Rowlands et al. 2014) have elucidated that cold diffuse emission is relevant mainly for SFRs $\psi\la 30\, M_\odot$ yr$^{-1}$, but becomes less and less important at higher SFRs $\psi\ga 10^2\, M_\odot$ yr$^{-1}$. The same conclusion holds for high redshift $z\sim 1.5-3$ starforming galaxies with SFR $\psi\ga 30\, M_\odot$ yr$^{-1}$, as it emerges from the analysis of the ALESS survey by Swinbank et al. (2014) and da Cunha et al. (2015, see their Fig. 10), who find dust temperatures in excess of $30$ K.

When basing solely on UV measurements, the main concern is to correct for
dust extinction. One of the most common method is to exploit the correlation
between the UV slope $\beta_{\rm UV}$ and the IRX ratio as gauged in the
local Universe (e.g., Meurer et al. 1999; Reddy et al. 2012; Bouwens et al.
2015). However, for SFRs $\psi\ga 30\, M_\odot$ yr$^{-1}$ when the
attenuation becomes appreciable, the $\beta_{\rm UV}$-IRX correlation is
found to be extremely dispersed, resulting in a very uncertain estimate of
the attenuation even in local samples (e.g., Howell et al. 2010; Reddy et al. 2015). On the other hand, the correlation is found to be less scattered for
SFRs $\psi\la 30\, M_{\odot}$ yr$^{-1}$, and the dust correction to UV
luminosity gets more secure and relatively small on the average. This is also suggested by the UV attenuation inferred by combining H$\alpha$ measurements with the Calzetti extinction curve (e.g., Mancuso et al. 2015; Reddy et al. 2015).

\begin{figure}
\epsscale{1}\plotone{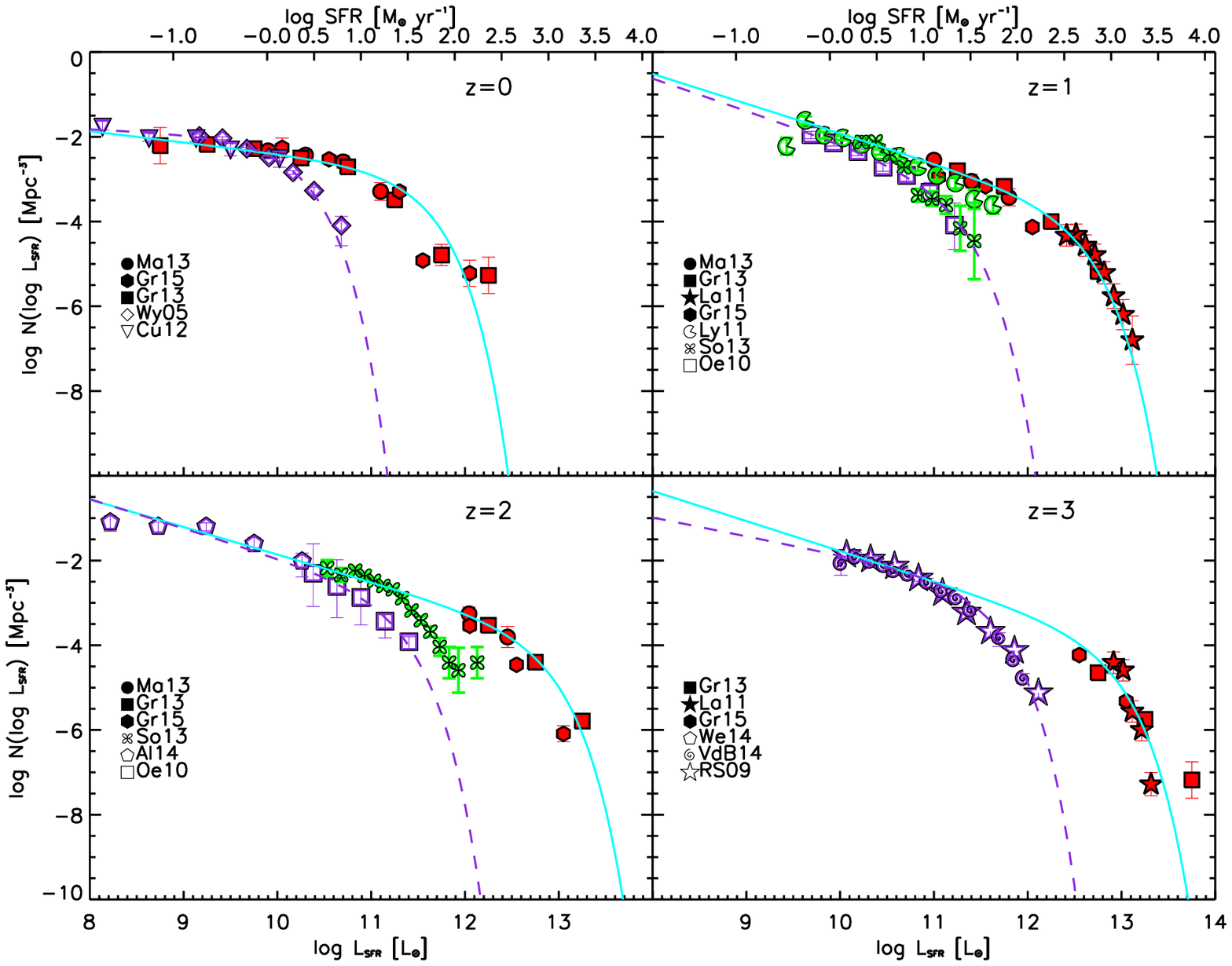}\caption{The SFR function
at redshifts $z\approx 0-3$. Solid cyan lines illustrate our fits to the
intrinsic (IR+UV) SFR functions, while violet dashed lines refer to the UV-inferred SFR functions. UV data (dust-corrected; violet symbols) are from Wyder et al. (2005; open diamonds), Cucciati et al. (2012; open inverse triangles), Oesch et al. (2010; open squares), Alavi et al. (2014; open pentagons), Reddy \& Steidel (2009; open stars), van der Burg et al. (2010; spirals), H$\alpha$ data (green symbols) from Ly et al. (2011; pacmans), Sobral et al. (2013; clovers), and IR data (red symbols) from Magnelli et al. (2013; filled circles), Gruppioni et al. (2013; filled squares), Gruppioni et al. (2015; filled hexagons), Lapi et al. (2011; filled stars).}\label{fig|SFRlowz}
\end{figure}

Given that, we build up the intrinsic SFR function $N(\log \psi,z)$ as
follows. We start from the most recent determinations of the luminosity
functions at different redshifts from far-IR and UV data (the latter being dust-corrected according to the $\beta_{\rm UV}$-IRX relation, see Meurer et al. 1999; Bouwens et al. 2009, 2015); the outcome is illustrated in
Fig.~\ref{fig|SFRlowz}. The SFR $\psi$ and the associated luminosity
$L_{\psi}$ reported on the upper and lower axis have been related assuming
the calibration
\begin{equation}\label{eq|Lsfr}
\log {\psi\over M_\odot~{\rm yr}^{-1}} \approx -9.8+\log {L_{\psi}\over
L_\odot}~,
\end{equation}
approximately holding for a Chabrier's IMF both for far-IR and (intrinsic) UV luminosities (see Kennicutt \& Evans 2012).

At redshift $z\la 3$, we lack a robust determination of the SFR function at
intermediate values of the SFR. On the one hand, UV data almost disappear for SFRs $\psi\ga 30\, M_\odot$ yr$^{-1}$ because of dust extinction (with
dust corrections becoming progressively uncertain, as discussed above). On
the other hand, far-IR data progressively disappear for SFRs $\la 10^{2}\,
M_\odot$ yr$^{-1}$ because of current observational limits. At higher
redshift $z\ga 4$, once more UV surveys can afford reliable estimate of the
SFR function for SFRs $\psi\la 30\, M_\odot$ yr$^{-1}$, but we lack far-IR
data deep enough to statistically probe the high-SFR end.

To obtain an analytic rendition of the intrinsic SFR function in the full
range of SFRs $\psi\sim 10^{-1}-$ several $10^3\, M_\odot$ yr$^{-1}$ and
redshift $z\sim 0-10$, we perform a least $\chi^2$-fit to the data with a
standard Schechter functional shape
\begin{equation}
N(\log\psi) = \mathcal{N}(z)\, \left[\psi\over\psi_c(z)\right]^{1-\alpha(z)}\,e^{-\psi/\psi_c(z)}~.
\end{equation}
The fit is educated, meaning that for redshift $z\la 3$, where both UV and
far-IR data are present, we consider as reliable the UV data (dust-corrected
according the $\beta_{\rm UV}$-IRX ratio) for SFRs $\psi\la 30\, M_{\odot}$
yr$^{-1}$, and the far-IR data for SFRs $\psi\ga 10^2\, M_{\odot}$ yr$^{-1}$. As for higher redshift, we make the assumption that at $z\ga 8$ the
(dust-corrected) UV data are reliable estimators of the intrinsic SFR
function. This assumption relies on the fact that with an age of the Universe shorter than $6\times 10^8$ yr, the amount of dust in a star-forming galaxy is expected to be rather small (see \S~\ref{sec|UVnature}).

Equipped with such values of the Schechter parameters at redshift $z\la 3$
and $z\ga 8$, we fit their evolution with a polynomial in log-redshift; in
other words, for any parameter $p(z)$ of the Schechter function, say
$\mathcal{N}(z)$, $\psi_c(z)$, or $\alpha(z)$, we fit for the functional
shape
\begin{equation}
p(z) = p_0 + p_1\, \xi + p_2\, \xi^2 + p_3\, \xi^3~,
\end{equation}
where $\xi\equiv\log(1+z)$. The outcomes of the fits are reported in Table~1.

\begin{figure}
\epsscale{1}\plotone{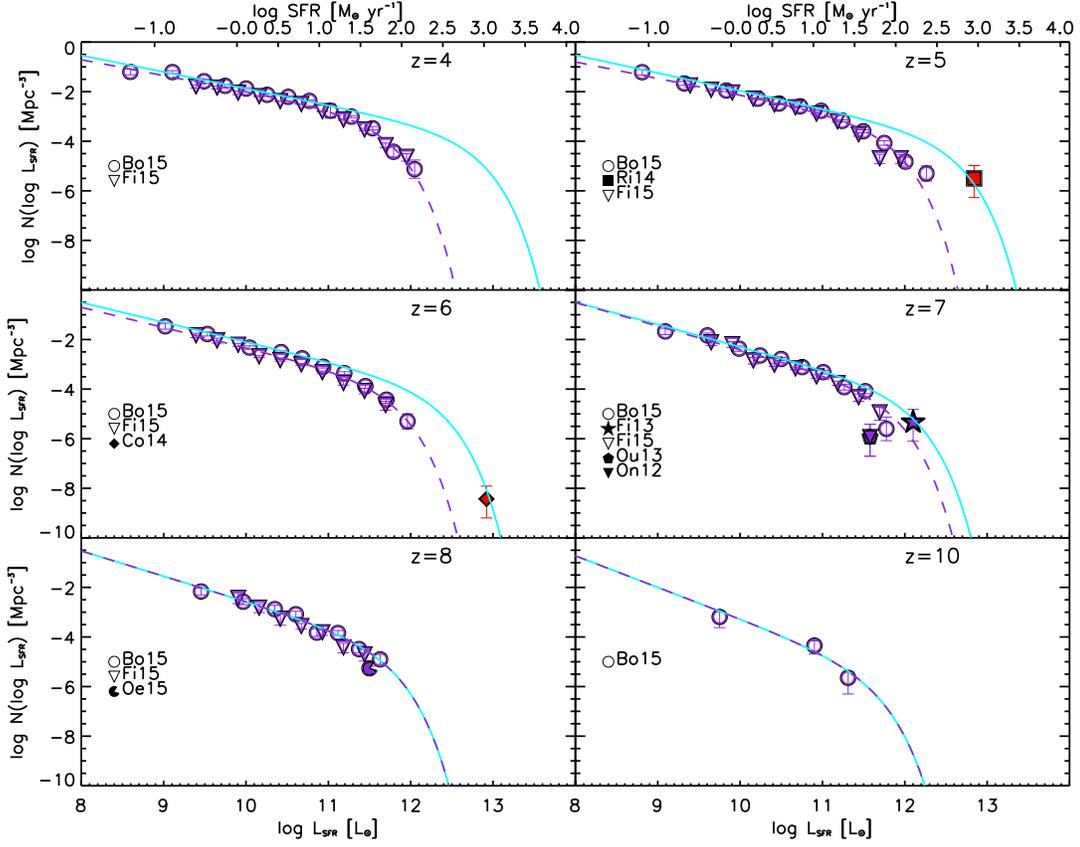}\caption{Same as previous figure but for redshifts $z\approx 4-10$. Solid cyan lines illustrate our
determination of the intrinsic (IR+UV) SFR functions, while violet dashed
lines illustrate our fits to the UV-inferred SFR function. UV data (dust-corrected; violet symbols) are from Bouwens et al. (2015; open
circles) and Finkelstein et al. (2015a; open inverse triangles). Filled
symbols represent the number density associated to the detection of
individual galaxies with spectroscopic redshift determination (see text for
details): violet ones refer to galaxies selected in UV/Ly$\alpha$ from
Finkelstein et al. (2013; star), Ouchi et al. (2013; pentagon), Ono et al.
(2012; inverse triangle), Oesch et al. (2015; pacman); red ones refer to
galaxies selected in IR/sub-mm from Riechers et al. (2014; square), Cooray et al. (2014; diamond).}\label{fig|SFRhighz}
\end{figure}

This procedure, based on the assumption of analytical continuity of the
intrinsic SFR function, yields a rendition that works pleasingly well for
$z\la 3$ and $z\ga 8$, see Figs.~\ref{fig|SFRlowz} and \ref{fig|SFRhighz};
moreover, we end up with an estimate for the behavior of the SFR function at
$z\sim 4-8$ where sampling by far-IR surveys is absent. In such a redshift
range, this estimate implies a significant number density of dusty starforming galaxies with SFR $\psi\ga 10^2\, M_{\odot}$ yr$^{-1}$, currently missed by UV data (even corrected for dust extinction). To highlight more clearly this point, we also report in Figs.~\ref{fig|SFRlowz} and \ref{fig|SFRhighz} the SFR function that would have been inferred basing solely on dust-corrected UV data over the full redshift range $z\approx 0-10$. Plainly, at any redshift $z\la 7$ UV data, even corrected for dust extinction basing on the UV slope, strongly underestimate the intrinsic SFR function for SFRs $\psi\ga 30\, M_\odot$ yr$^{-1}$.

Circumstantial evidence for such a population of dusty star-forming galaxies
at $z\ga 4$ is accumulating over the recent years. Riechers et al. (2014)
detected a dust obscured galaxy at $z\approx 5.3$ with SFR $\psi\approx
1100\,M_{\odot}$ yr$^{-1}$ and stellar mass $M_\star\approx 10^{10}\,
M_\odot$. Cooray et al. (2014) detected a second one at $z\approx 6.34$ with
SFR $\psi\approx 1320\,M_{\odot}$ yr$^{-1}$ and stellar mass $M_\star\approx
5 \times 10^{10}\, M_\odot$. It is remarkable that the inferred number
densities for these objects, though within the considerable uncertainties,
agree with the prediction of our intrinsic SFR function, while being
substantially higher than the expectations from the purely UV-inferred one
(see Fig.~\ref{fig|SFRhighz}).

At lower levels of SFRs, moderately dusty galaxies start to be selected even in the UV, especially at high redshift $z\ga 7$. For example, Finkelstein et al. (2013) detected one at $z\approx 7.51$ with SFR $\psi\approx 200\, M_{\odot}$ yr$^{-1}$ and stellar mass $M_\star\approx 6\times 10^8\, M_\odot$. Oesch et al. (2015) revealed one at $z\approx 7.73$
with SFR $\psi\approx 30-50\, M_{\odot}$ yr$^{-1}$ and stellar mass
$M_\star\approx 5\times 10^9\, M_\odot$. Ouchi et al. (2013) detected another one at $z\approx 6.6$ with SFR $\psi\approx 60\, M_{\odot}$ yr$^{-1}$ and
stellar mass $M_\star\approx 10^{10}\, M_\odot$. Ono et al. (2012) detected
one at $z\approx 7.2$ with SFR $\psi\approx 60\, M_{\odot}$ yr$^{-1}$ and
stellar mass $M_\star\approx 3\times 10^8\, M_\odot$. Intense search is
currently going on, with an appreciable number of candidates being found (see Roberts-Borsani et al. 2015; Zitrin et al. 2015). The number densities of
these galaxies are consistent with the UV-corrected SFR function, that at
these high redshift approaches the intrinsic one.

We stress that while the focus of the present paper is mainly on the bright
portion of the intrinsic SFR function at high-redshift, the faint end as
sampled by UV data is essential to understand important issues both in
astrophysics/cosmology like the history of cosmic reionization (e.g., Cai et
al. 2014; Robertson et al. 2015) and even in fundamental physics like the
nature of dark matter (e.g., Lapi \& Danese 2015).

\subsection{Validating the intrinsic SFR function via the (sub-)mm
counts}\label{sec|counts}

We aim at validating our intrinsic SFR function via comparison with the
observed (sub-)mm counts, redshift distributions, and cosmic infrared background. We compute the counts according to the expression
(see Lapi et al. 2011)
\begin{equation}
{{\rm d}N\over {\rm d}\log S_\nu\,{\rm d}\Omega}(S_\nu) = \int{\rm
d}z~{{\rm d}V\over {\rm d}z\,{\rm d}\Omega}~N(\log\psi)\,{{\rm
d}\log\psi\over {\rm d}\log S_\nu}~
\end{equation}
in terms of the flux
\begin{equation}
S_\nu = {L_{\nu\,(1+z)}\over L_\psi}\,{(1+z)\over 4\pi\,D_L^2(z)}~;
\end{equation}
in the above $N(\log\psi)$ is the SFR function, ${\rm d}V/{\rm
d}z\,{\rm d}\Omega$ is the cosmological volume per redshift bin and unit
solid angle, $L_\psi$ is the bolometric luminosity associated to the SFR
$\psi$ according to Eq.~(\ref{eq|Lsfr}), and $L_{\nu\,(1+z)}/L_\psi$ is
the $K-$correction. The latter has been computed basing on the spectral
energy distribution (SED) typical of a high-redshift, dust-obscured
star-forming galaxy; specifically, we consider as a reference the SED of the
`Cosmic Eyelash' (SMM J2135+0102; see Swinbank et al. 2010; Ivison et al.
2010), but we shall show the impact of assuming a different SED. Actually, for the sources located at $z\la 0.3$ and contributing only
to the very bright counts probed by \textsl{Planck Collaboration VII} (2013), we have adopted the warmest SED from the template library by Smith et al.
(2012).

We have also evaluated the contribution to the counts from strong
galaxy-scale gravitational lensing, according to the SISSA model (cf. Lapi et al. 2012); the lensed counts are computed as
\begin{eqnarray}
\nonumber{{\rm d}N_{\rm lensed}\over {\rm d}\log S_\nu\,{\rm d}\Omega}(S_\nu) && = \int{\rm d}z\,{1\over
\langle\mu\rangle}\,\int_2^{\mu_{\rm max}}{\rm d}\mu~{{\rm d}p\over {\rm d}\mu}(\mu,z)\, \times\\
\\
\nonumber && \times {{\rm d}N_{\rm unlensed}\over {\rm d}\log S_\nu\,{\rm
d}\Omega\, {\rm d}z}(S_\nu/\mu,z)~.
\end{eqnarray}
Here ${\rm d}p/{\rm d}\mu$ is the amplification distribution and
$\langle\mu\rangle$ is its average ($\approx 1$ for a wide-area survey); a
maximum amplification of $\mu_{\rm max}\approx 25$ has been adopted (see Cai
et al. 2013; Bonato et al. 2014).

\begin{figure}
\epsscale{1}\plottwo{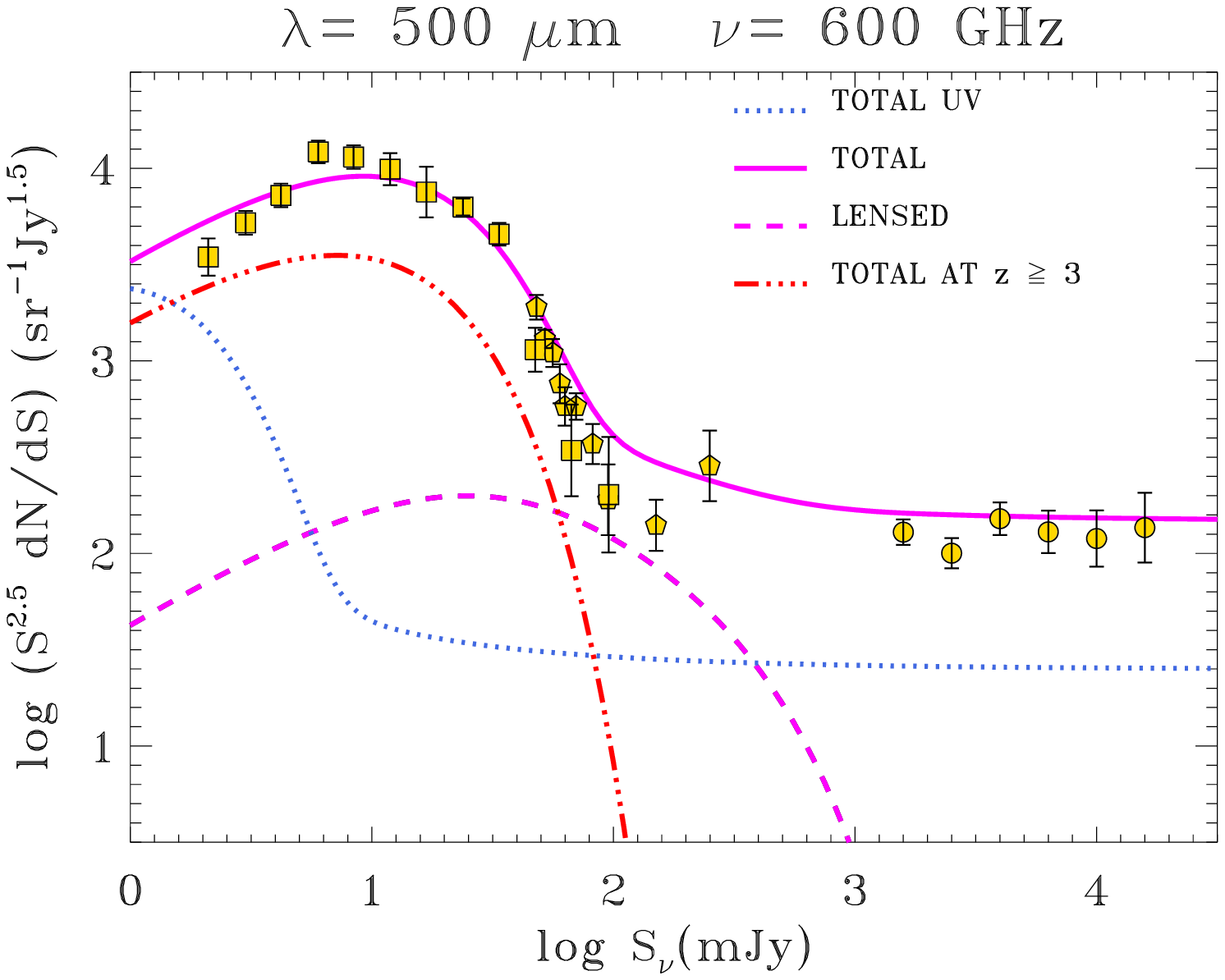}{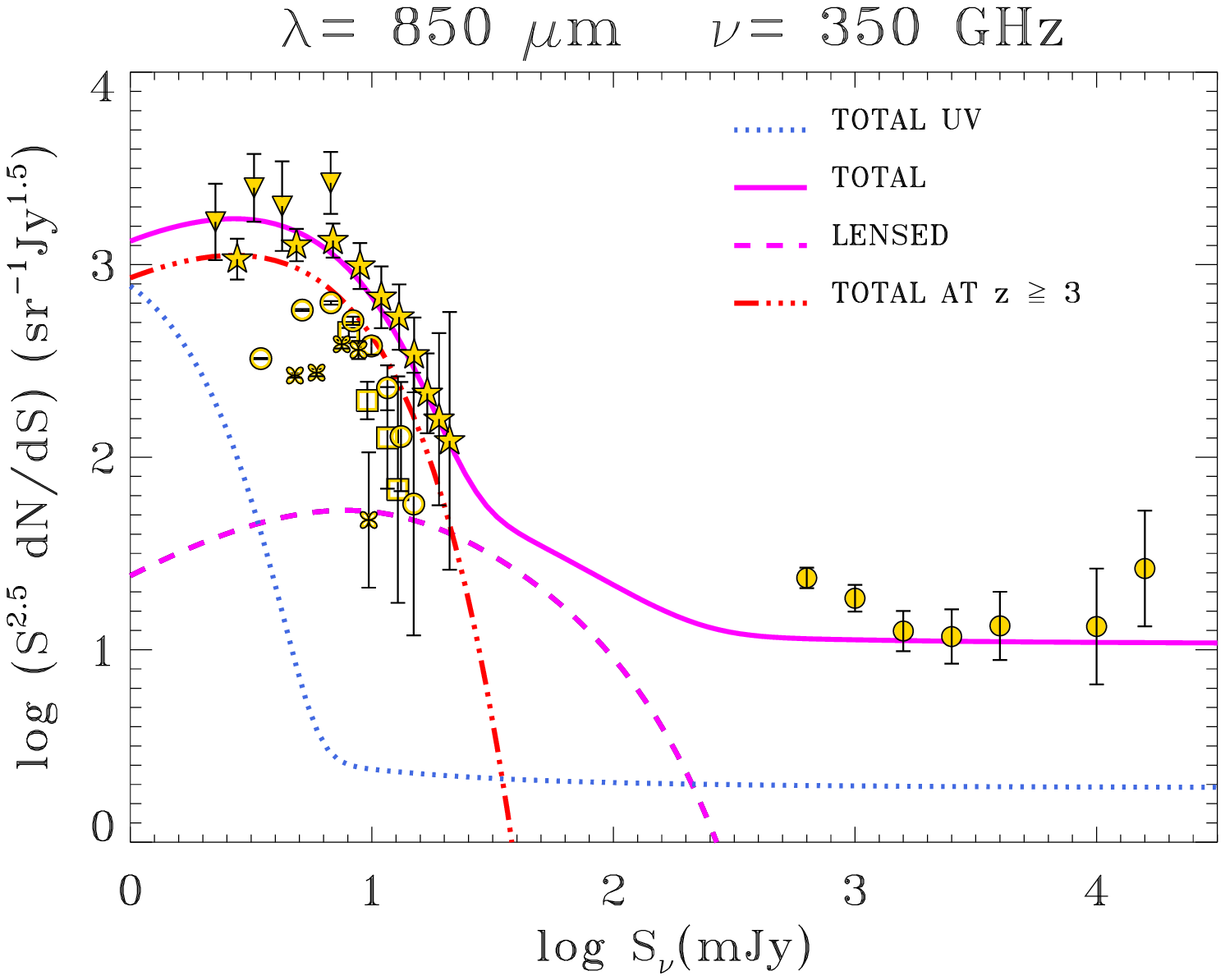}\\
\plottwo{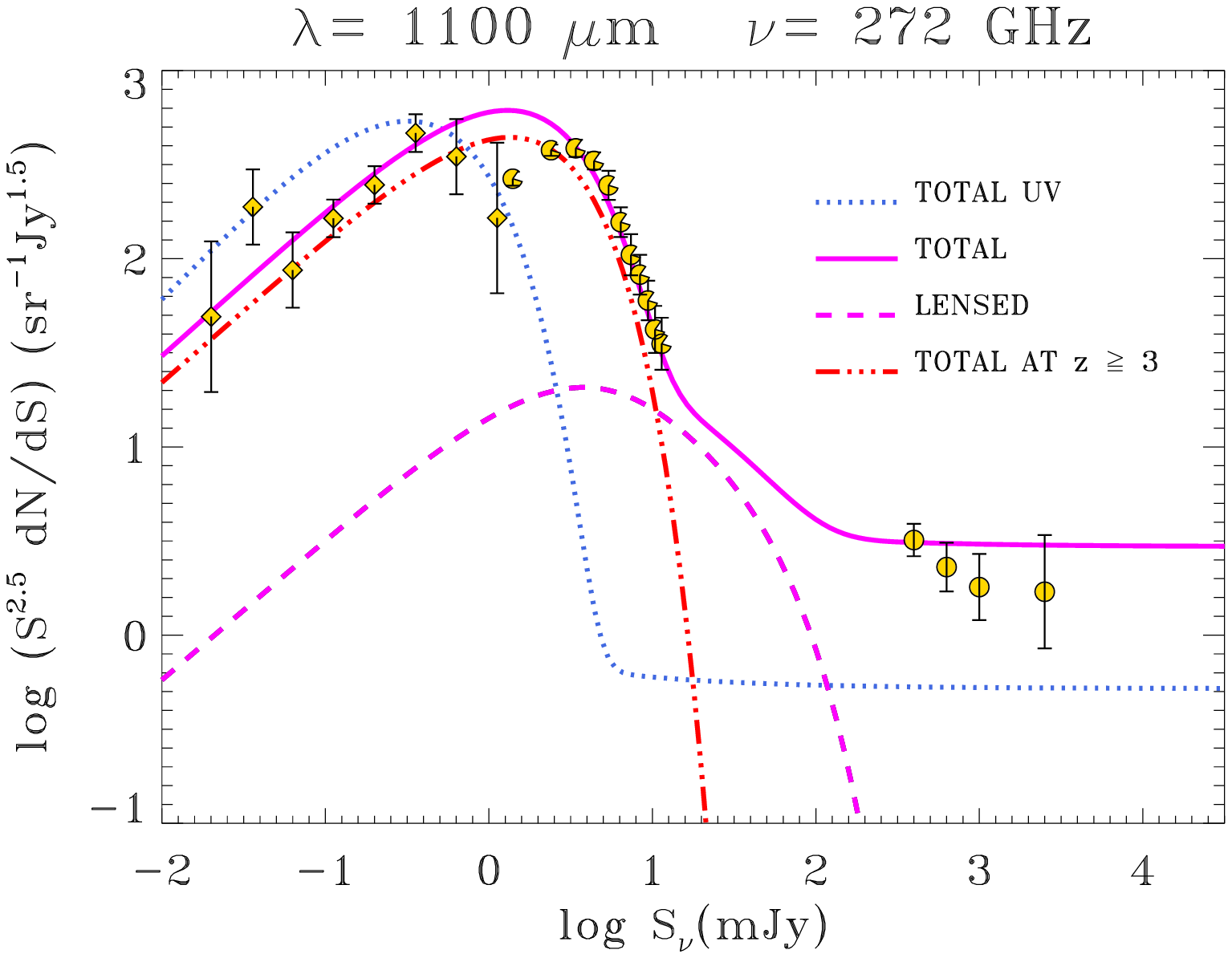}{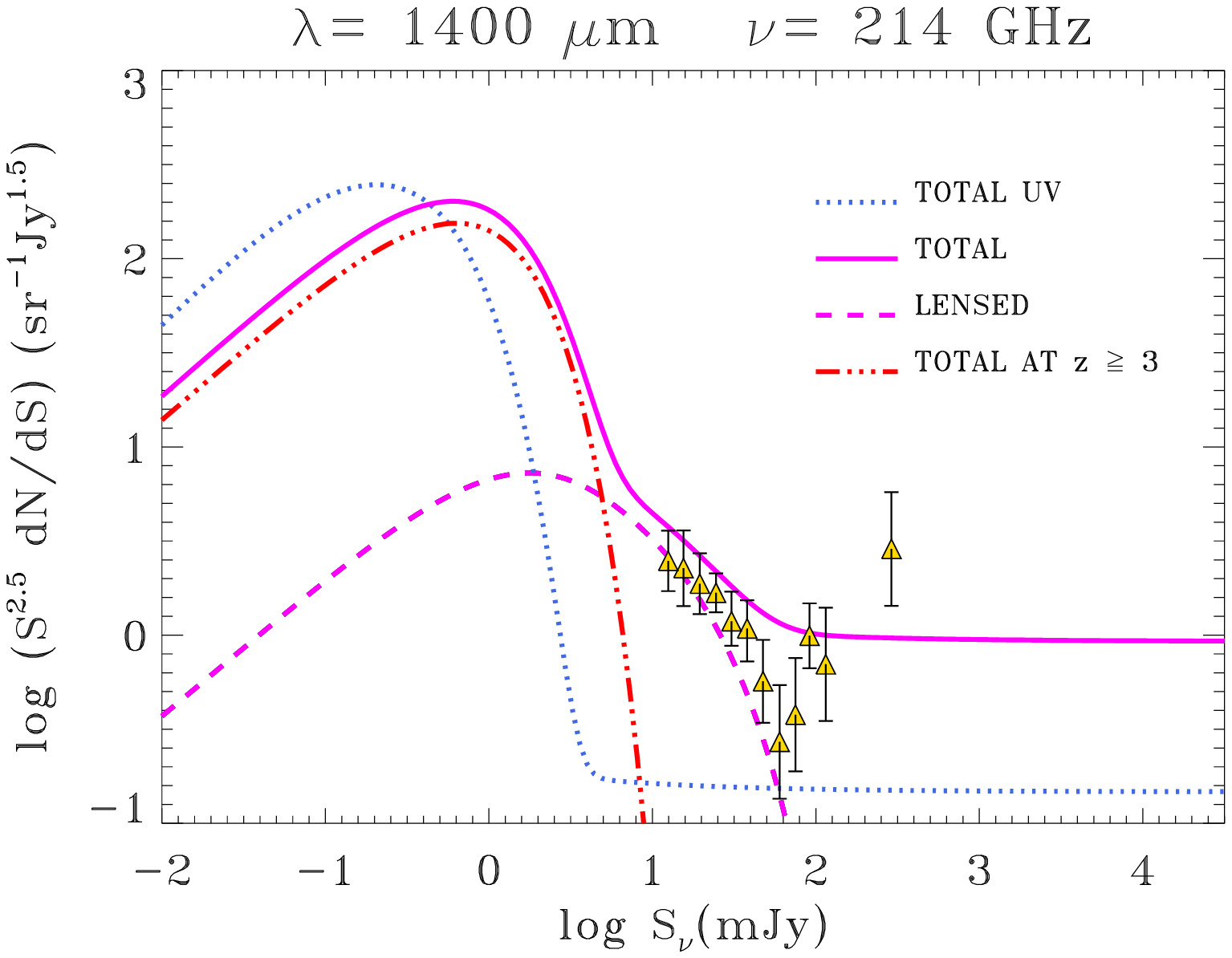}
\caption{Euclidean-normalized differential number counts at $500$ (top left), $850$ (top right), $1100$ (bottom left), and $1400\, \mu$m (bottom right). Magenta lines refer to the counts derived from our intrinsic SFR function; the contribution to the total counts (solid) from strongly lensed galaxies (dashed) is highlighted. The triple dot-dashed red line is the contribution to unlensed counts from galaxies at $z\ga 3$. The blue dotted line refers to the counts derived from the UV-inferred SFR function. Data (gold symbols) are from \textsl{Planck} collaboration (2013; filled circles), \textsl{Herschel}/HerMES by Bethermin et al. (2012; filled squares), \textsl{Herschel}/ATLAS by Clements et al. (2010; filled pentagons), \textsl{SCUBA} by Coppin et al. (2006; filled stars) and by Noble et al. (2012; filled reversed triangles), \textsl{LABOCA} by Weiss et al. (2009; open circles), \textsl{ALMA} by Karim et al. (2013; open clovers), Simpson et al. (2015; open squares) and Fujimoto et al. (2016; filled diamonds), \textsl{AzTEC} by Scott et al. (2012; filled pacmans), and \textsl{SPT} by Mocanu et al. (2013; filled triangles).}\label{fig|diffcounts}
\end{figure}

The Euclidean-normalized, differential counts at various wavelengths
$\lambda\approx 500$, $850$, $1100$, and $1400\, \mu$m are plotted in
Fig.~\ref{fig|diffcounts}. We find an excellent agreement of the counts
derived from our intrinsic SFR function with various observational data (see details in the caption). By contrast, we also show that the counts expected from the UV-inferred SFR function considerably underpredict the data at the bright end. Note that, to make the contribution to the counts from the UV-inferred SFR function as large as possible, we have assumed that all the UV emission is reprocessed by dust and reradiated in the far-IR according to the coldest SED from the template library by Smith et al. (2012).

\begin{figure}
\epsscale{0.7}\plotone{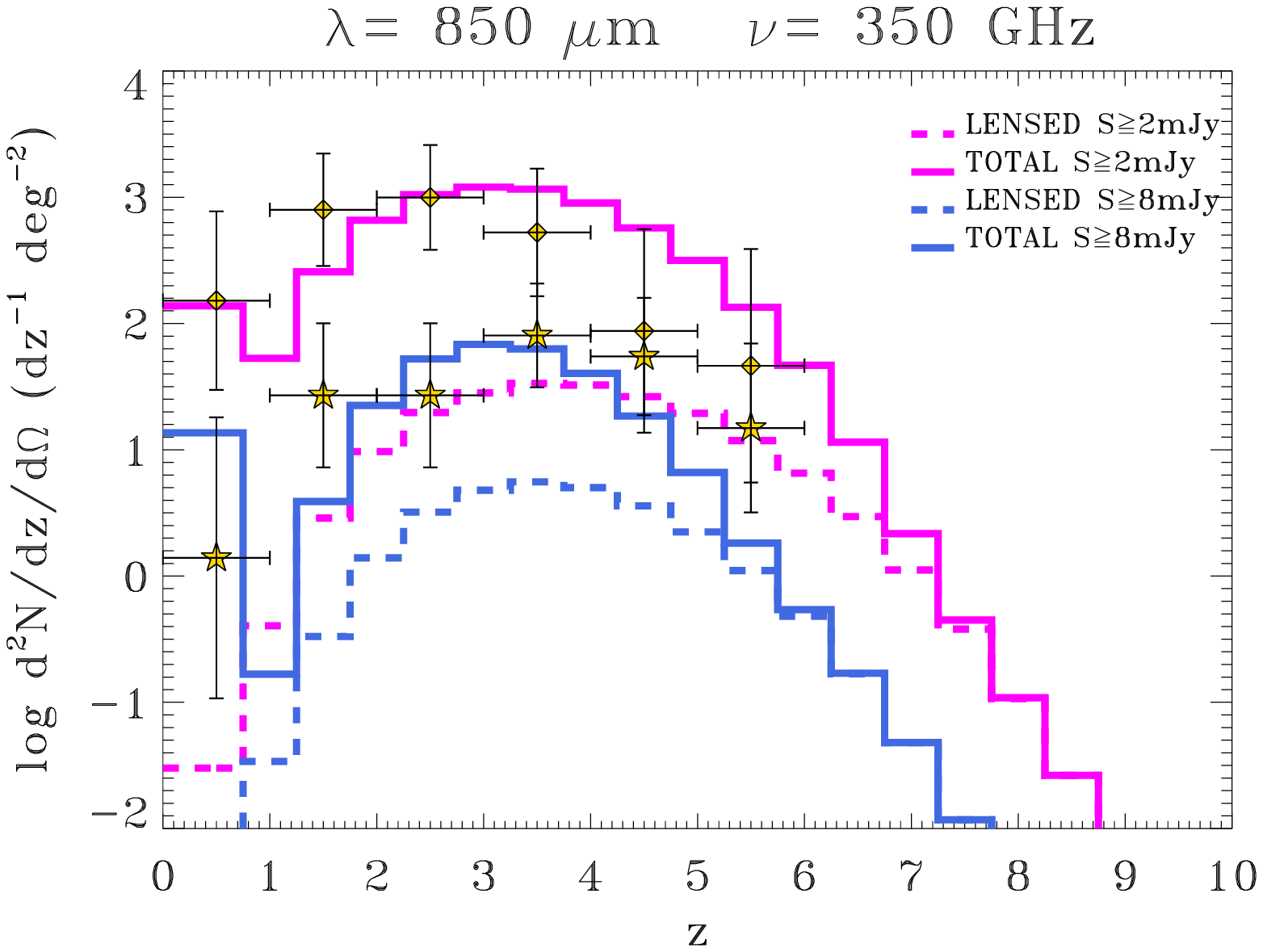}\\\plotone{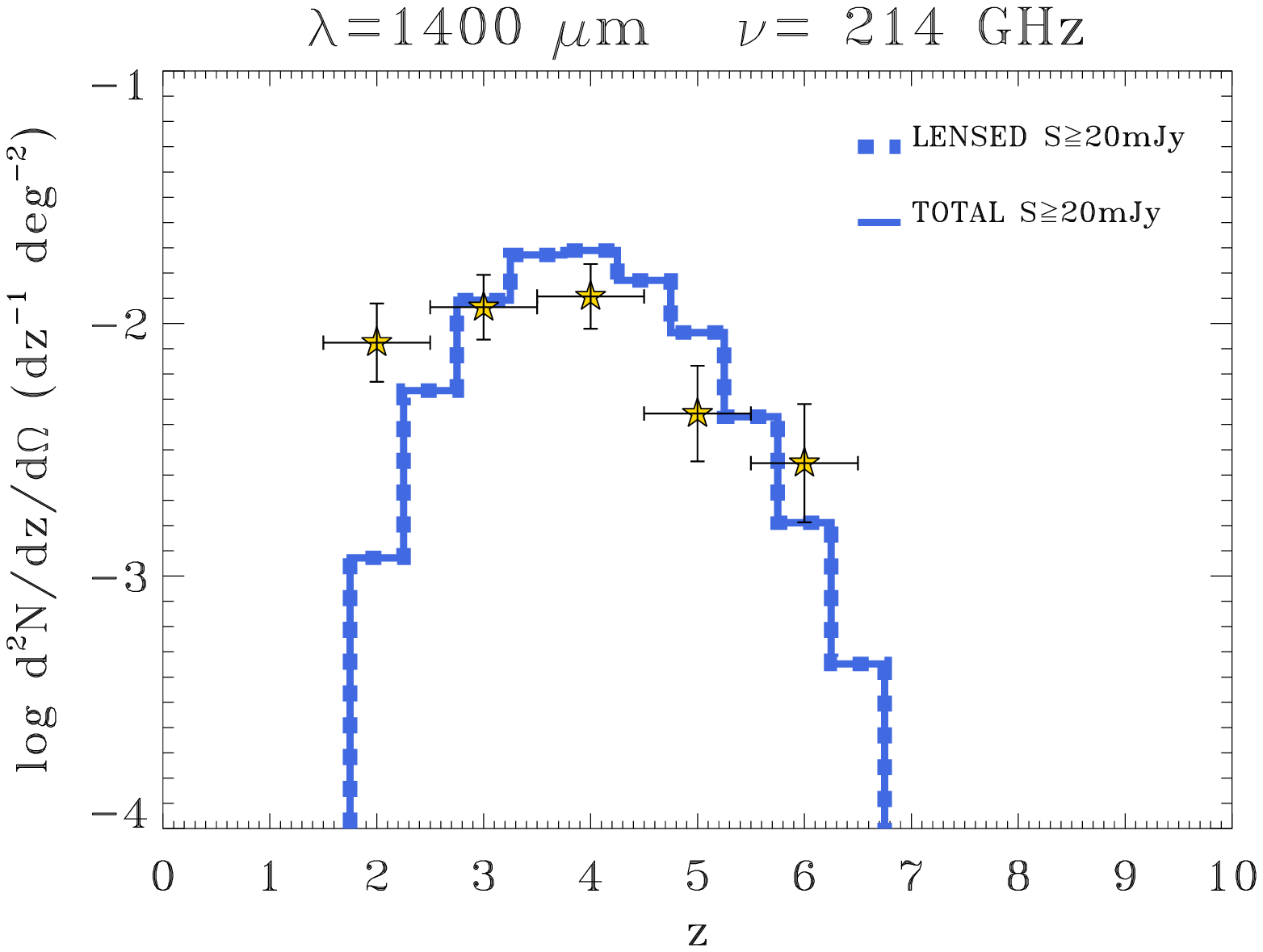}
\caption{Redshift distributions at $850$ (top panel) and $1400\, \mu$m (bottom panel). At $850\, \mu$m magenta lines refer to a limiting flux of $2$ mJy and blue lines to a limiting flux of $8$ mJy, with the contribution to the total (solid) from strong galaxy-scale
gravitational lensing (dashed) highlighted; data are from \textsl{AzTEC}-\textsl{LABOCA} by Koprowski et al. (2014, stars), and from \textsl{SCUBA-2} by Koprowski et al. (2015, diamonds). At $1400\, \mu$m blue lines (solid and dashed are superimposed) refer to a limiting flux of $20$ mJy; data are from \textsl{ALMA}-\textsl{SPT} by Weiss et al. (2013, stars).}\label{fig|zdistdata}
\end{figure}

We remark that the counts at $\lambda\ga 850\, \mu$m for fluxes $\ga$ several mJy are substantially contributed by galaxies located at $z\ga 3$. This is shown in detail by the redshift distributions presented in Fig.~\ref{fig|zdistdata}, that peak at $z\approx 3-4$ with a substantial tail at higher $z$. Specifically, we find a pleasing agreement of our results based on the intrinsic SFR function with the $1400\,\mu$m \textsl{ALMA}-\textsl{SPT} data at a flux limit of $\ga 20$ mJy, constructed from a sample of $26$ galaxies with spectroscopic redshifts (Weiss et al. 2013). The redshift distribution is essentially contributed by gravitationally lensed sources, and it constitutes an extremely important test of the intrinsic SFR function up to $z\approx 6$. Note that the lensed counts would be strongly underestimated when basing on the (dust-corrected) UV-inferred SFR function (cf. Fig.~\ref{fig|diffcounts}, bottom right panel). We also find good agreement with the $850\,\mu$m data from \textsl{SCUBA-2} by Koprowski et al. (2015) at a limiting flux of $\ga 2$ mJy, from \textsl{AzTEC}-\textsl{LABOCA} data by Koprowski et al. (2014; see also Smolcic et al. 2012) at a flux limit of $\ga 8$ mJy, that constitute a sample of about $100$ sources with mostly photometric redshifts.

\begin{figure}
\epsscale{1.0}\plotone{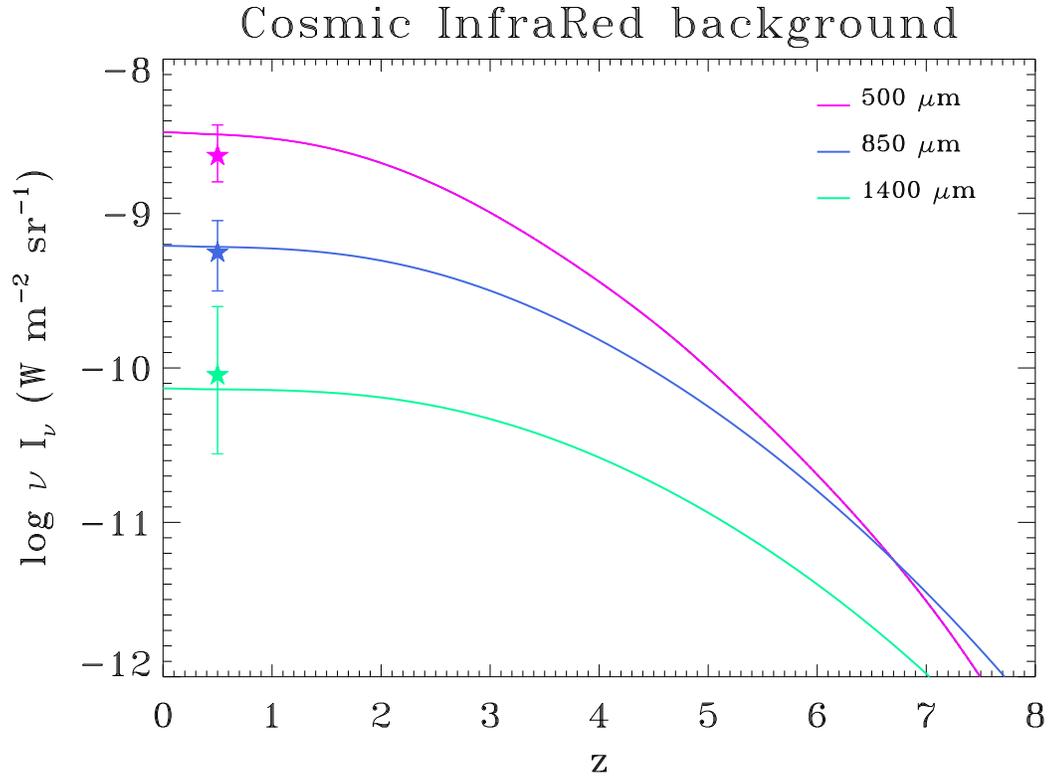}
\caption{Contribution to the the cosmic infrared background at $500$ (magenta), $850$ (blue), and $1400\, \mu$m (green) from redshift greater than $z$, as derived from our intrinsic SFR function, compared with the observational determinations at $z\approx 0$ (stars, slightly offset in redshift for clarity) by Fixsen et al. (1998; see also Lagache et al. 1999).}\label{fig|back}
\end{figure}

In Fig.~\ref{fig|back} we show that the cosmic infrared background at $500$, $850$, and $1400\, \mu$m as derived from our intrinsic SFR function is consistent with the measurements by Fixsen et al. (1998; see also Lagache et al. 1999; Planck Collaboration XVIII 2011, XXX 2014). As extensively discussed by Lapi et al. (2011; see their Fig.~19), the evolution with redshift of the background highlights that for $\lambda\ga 500\, \mu$m it is mostly contributed by high-redshift galaxies down to $z\approx 2-3$. This trend strengthens as $\lambda$ increases; in particular, at $1400\, \mu$m about $50\%$ of the background is contributed by dusty galaxies at $z\ga 3$; this fraction would drop dramatically to less than $10\%$ basing on the (dust-corrected) UV-inferred SFR function.

All in all, the agreement with the observed counts, redshift distributions (including lensed sources), and cosmic infrared background constitutes a robust validation of our intrinsic SFR function in a range of SFRs and redshift where the far-IR data on the luminosity function are still not available.

\begin{figure}
\epsscale{0.7}\plotone{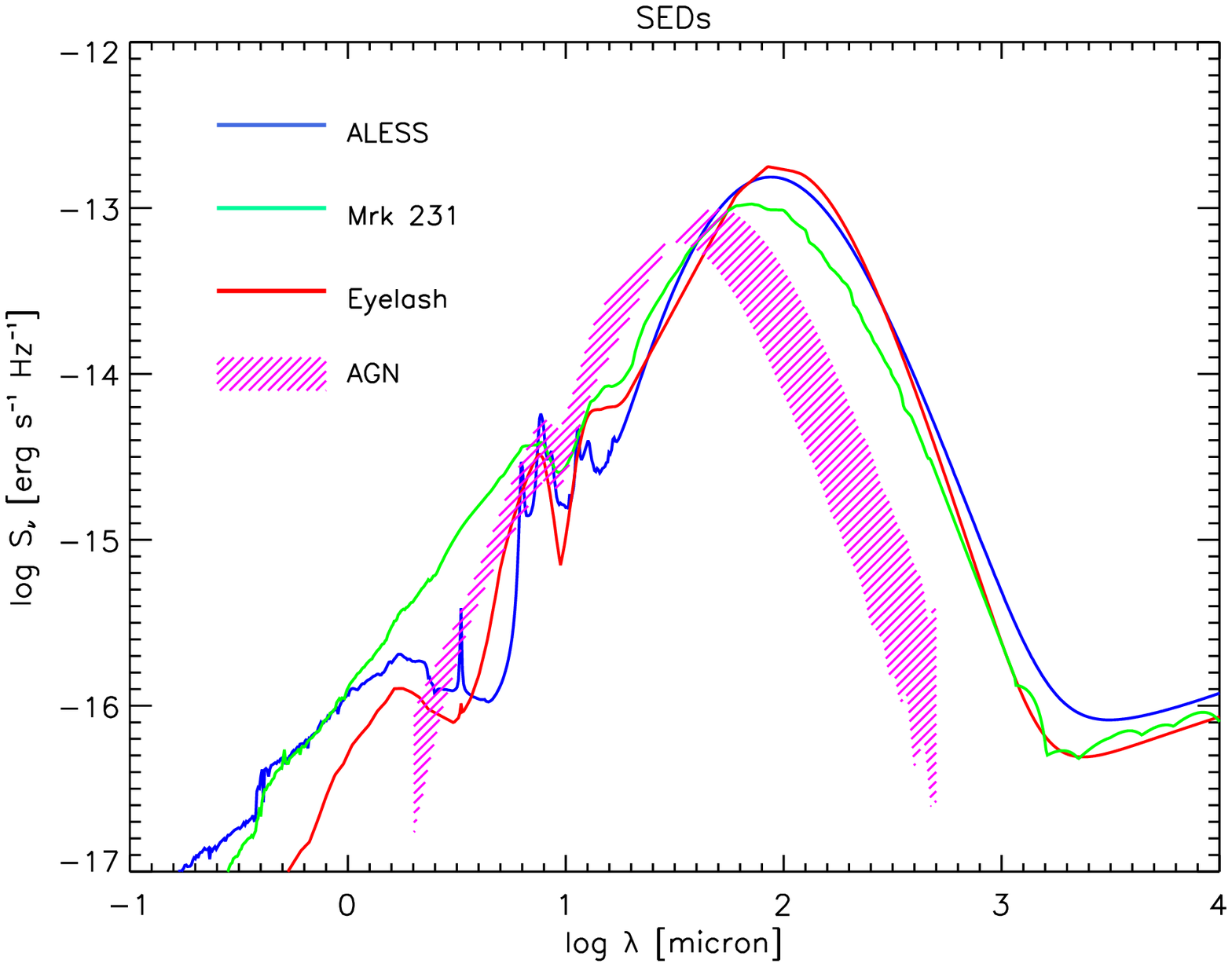}\\\plotone{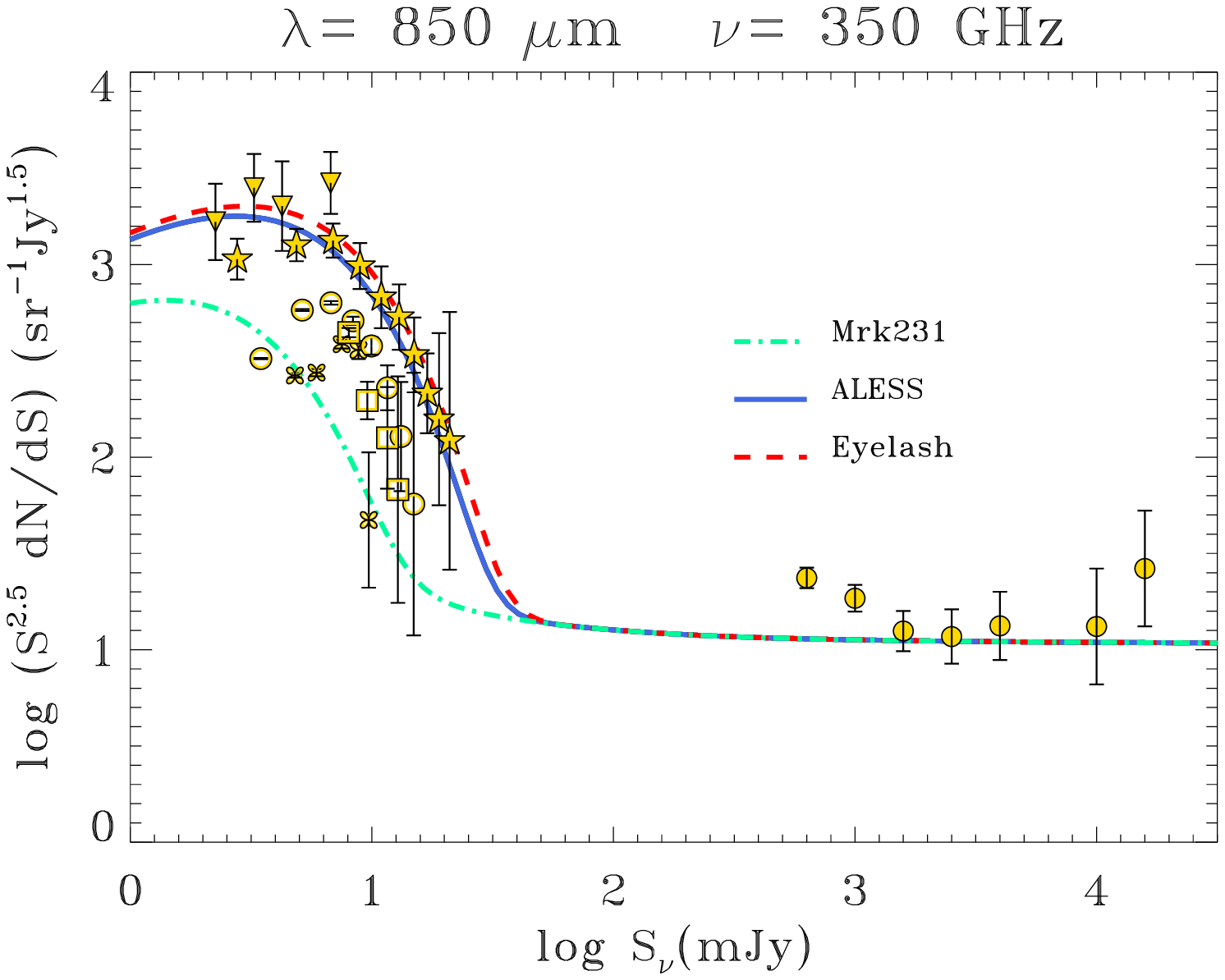}
\caption{Top panel: comparison among the SEDs of three typical dust-obscured, star-forming galaxies (normalized in the range $3-1100\, \mu$m to $1$ erg s$^{-1}$): red line refers to the Cosmic Eyelash (Ivison et al. 2010), blue line to the average from ALESS galaxies (Swinbank et al. 2014; da Cunha et al. 2015), and green line to Mrk231 (e.g., Polletta et al. 2007). The typical SEDs of obscured AGNs (including both low- and high-$z$ objects) is plotted as a magenta region (Siebenmorgen et al. 2015). Bottom panel: effect of varying the SED on the total $850\,\mu$m counts; data points as in Fig.~\ref{fig|diffcounts}.}\label{fig|sedcomp}
\end{figure}

Two remarks are in order. First, we have investigated the impact of using different SEDs typical of star-forming, dust-obscured galaxies, namely, the Cosmic Eyelash (our reference, see above), the average from the ALESS sample (Swinbank et al. 2014; da Cunha et al. 2015; this is similar to the classic SED of Arp220, see Rangwala et al. 2011), and the local ULIRG+Seyfert1 galaxy Mrk231 (e.g., Polletta et al. 2007). In Fig.~\ref{fig|sedcomp} we show that the effect on the steep part of the $850\,\mu$m counts (mostly contributed by $z\ga 2-3$) is small when changing from the Eyelash to the ALESS (or Arp220) SED; the same holds at any $\lambda\ga 500\, \mu$m, since for galaxies at $z\ga 2-3$ the SEDs are quite similar in the corresponding range of rest-frame wavelengths. On the contrary, considering a SED shape like that of Mrk231, which exhibits more power in the mid-IR regime, would appreciably underpredict the (sub-)mm counts.

Second, we point out that the contribution of active galactic nuclei (AGNs) is marginally relevant as to the above statistics. As a matter of fact, the AGN contribution to the IR emission, generally ascribed to the presence of a dusty torus, is characterized by SEDs peaking at $20-40\,\mu$m with a rapid fall-off at longer wavelengths, because the emission is dominated by hot dust grains. Theoretical works have shown that this fall-off is weakly dependent on geometry, clumpiness of the torus, and orientation of the line of sight (e.g., Pier \& Krolik 1992; Granato \& Danese 1994; Efstathiou \& Rowan-Robinson 1995; Nenkova et al. 2008). The results of these works well match the observed SEDs of local AGNs.

In Fig.~\ref{fig|sedcomp} (top panel) we present the typical restframe SEDs of obscured AGNs (referring to both local and high$-z$ objects) as fitted by Siebenmorgen et al. (2015). This plot illustrates the relative contribution of an obscured AGN and of its host dusty galaxy, under the assumption that their integrated luminosity over the range $3-1100\, \mu$m is the same (a conservative hypothesis for most of (sub-)mm selected galaxies). In order to test the SFR function at $z\ga 3$, the statistics of counts and lensed objects at $\lambda\sim 1400\,\mu$m (Weiss et al. 2013) are extremely informative. This observational wavelength corresponds to restframe $\lambda\ga 200\,\mu$m for galaxies at $z\la 6$. From Fig.~\ref{fig|sedcomp} it is apparent that the obscured AGN flux/luminosity is $\la 10\%$ with respect to that of the host galaxy.

This conclusion is also supported by Delvecchio et al. (2014), who have performed broad-band SED decomposition in about $4000$ galaxies detected at $160\,\mu$m by \textsl{Herschel} in the redshift range $z\sim 0-3$. They find that the SEDs of galaxies with appreciable AGN contribution (around $30\%$ of the total) are well fit by the standard starburst component peaking at $\approx 100\,\mu$m and by an AGN component peaking at $20-40\, \mu$m (cf. Fig.~5 of Delvecchio et al. 2014). Moreover, the same approach have been exploited by Gruppioni et al. (2015) in the COSMOS and GOODS-S fields of the PEP and HerMES/\textsl{Herschel} surveys to subtract the AGN emission on an object-by-object basis, and then to reconstruct the bright end of the SFR function at $z\la 3$ (cf. hexagons in Fig.~\ref{fig|SFRlowz}); as a matter of fact, their outcomes agree with our SFR function, again indicating that the AGN contribution is irrelevant.

Swinbank et al. (2014) and da Cunha et al. (2015) reported observations with \textsl{ALMA} of $99$ high$-z$ sub-mm galaxies in the ECDFS with multi-wavelength observations, covering a very wide spectral range. These authors show that the composite spectrum of such galaxies in the range $\lambda \sim 10-1000\, \mu$m can be well represented by the superposition of three grey-bodies referring to different dust components, cold, warm and hot, with temperatures $T\approx 20-30$ K, $50-60$ K and $80-120$ K, respectively. The hot component peaking at around $30\, \mu$m suggests the presence, in a statistical sense, of an AGN contribution, which becomes irrelevant at $\lambda \ga 70-80\, \mu$m.

All in all, the combination of the short lifetime for luminous AGNs and of their SED makes the AGN component irrelevant as for the far-IR/(sub-)mm counts. A general discussion on the luminosity function and counts of AGN type 1, 2 and 3 (the latter being those growing at the center of star-forming galaxies at substantial redshift) at wavelengths ranging from UV to mm bands was presented by Cai et al. (2013). We stress that these findings do not exclude relationships between the star formation and the central black hole accretion history (e.g., Alexander \& Hickox 2012; Kormendy \& Ho 2013; Lapi et al. 2014; Aversa et al. 2015).

\subsection{Validating the intrinsic SFR function via the continuity
equation}\label{sec|cont}

We now turn to validate our intrinsic SFR function by exploiting the observed stellar mass function at $z\ga 4$. The SFR and stellar mass functions are
naturally related via the continuity equation, along the lines already
pursued for lower redshifts by Aversa et al. (2015; see also Leja et al.
2015). The continuity equation has been originally devised for connecting the AGN statistics to the demographics of both active and dormant supermassive
black holes; recently, it has been also applied with remarkable success to
link the evolution of the galaxy SFR function $N(\psi,t)$ to the stellar mass functions $N(M_\star,t)$ of active and passively-evolving galaxies across
cosmic times. We defer the reader to the paper by Aversa et al. (2015) for an extensive discussion of this approach; here we just recall its basic features and implement some improvements.

The continuity equation in integral formulation can be written as
\begin{equation}\label{eq|continuity}
N(\psi,t) = \int_0^\infty{\rm d}M_\star\, \left[\partial_t
N(M_\star,t)\right]\, {{\rm d}\tau\over{\rm d}\psi}(\psi|M_\star,t)~;
\end{equation}
here $t$ is the cosmological time corresponding to redshift $z$, $\tau$ is
internal galactic time (i.e., the time elapsed since the triggering of
significant star formation) and ${\rm d}\tau/{\rm d}\psi$ is the time spent
by a galaxy with current stellar mass $M_\star$ in the SFR range
$[\psi,\psi+{\rm d}\psi]$ given a star formation history
$\psi=\psi(\tau|M_\star,t)$. Since we are mainly interested in the
high-redshift $z\ga 4$ evolution of the mass function, we have neglected any
source term due to `dry' merging, i.e., events adding the whole mass content
in stars of merging galaxies without contributing significantly to luminosity associated with star formation.

As for the star formation history $\psi(\tau|M_\star,t)$, Aversa et al.
(2015) have considered the standard, time-honored assumptions of a
constant, or exponentially increasing/decreasing SFR. Here we follow the
indications emerging from recent studies of SED-modeling (e.g., Papovich et
al. 2011; Smit et al. 2012; Moustakas et al. 2013; Steinhardt et al. 2014)
for a slow, powerlaw increase of the SFR with a characteristic time
$\tau_\star$, in the form
\begin{equation}\label{eq|continuity2}
\psi(\tau|M_\star,t) = \psi_\star\, \left({\tau\over
\tau_\star}\right)^\kappa~~~~~~~\psi_\star={M_\star\, (\kappa+1)\over
\tau_\star}~,
\end{equation}
with $\kappa\approx 0.5$; the second equation above just links the
normalization $\psi_\star$ of the SFR history to the current stellar mass $M_\star$. However, we checked that our results do not depend on this specific representation. The quantity ${\rm d}\tau/{\rm d}\psi$ entering the
continuity equation reads
\begin{eqnarray}\label{eq|continuity3}
\nonumber{{\rm d}\tau\over{\rm d}\psi}(\psi|M_\star,t) && =
{\tau_\star^{1+1/\kappa}\over\kappa}\, {\psi^{-1+1/\kappa}\over
[(1+\kappa)\,M_\star]^{1/\kappa}}\times\\
\\
\nonumber && \times\,\Theta_{\rm H}\left[\psi\le {(1+\kappa)\,
M_\star\over \tau_\star}\right]~;
\end{eqnarray}
the Heaviside function $\Theta_{\rm H}[\cdot]$ specifies that a galaxy with
current mass $M_\star$ cannot have shone at a SFR $\psi$ exceeding $M_\star\, (\kappa+1)/\tau_\star$.

At the high redshifts $z\ga 4$, of interest here, the stellar mass function is dominated by actively star-forming galaxies; thus we adopt the star-formation timescale $\tau_\star$ inferred from the observed main-sequence $\psi-M_\star$ (e.g., Rodighiero et al. 2011, 2014; Whitaker et al. 2014; Renzini \& Peng 2015; Speagle et al. 2014). Such a timescale
$\tau_\star=\tau_\star(\psi,t)$ is itself a function of the SFR/stellar mass
and cosmic time.

We exploit the determination of the main sequence by Speagle et al. (2014), which takes into account many samples with different primary selections (UV, optical, far-IR; cf. their Table 3). This is a good representation of the statistical average relationships between SFR and stellar mass for galaxies over their lifetime (see also Koprowski et al. 2015). Note that in the Speagle et al. determination, 'off-main sequence' galaxies are accounted for by a scatter of $0.3$ dex around the median relation, which is in turn dependent on redshift (see also Munoz \& Peeples 2015; da Cunha et al. 2015).

We point out that at lower redshifts $z\la 1.5$ it would be important to take the fraction of passively-evolving galaxies into account for obtaining sound estimates of the relic stellar mass function from the continuity equation (see Aversa et al. 2015; Leja et al. 2015). We also stress that $\tau_\star$ is in general different from the total duration of the star-formation episode over which most of the stellar mass is accumulated. More in detail, the two timescales are both quite close to a few $10^8$ yr for massive galaxies, which typically formed their stars in a violent burst with SFR $\psi\ga 10^2\, M_\odot$ yr$^{-1}$; but they can be appreciably different for less massive objects, which typically formed steadily their stars at much lower rates $\psi\la 10\, M_\odot$ yr$^{-1}$ over several Gyrs. Thus the total burst duration is an inverse function of the stellar mass in accord with the standard downsizing picture (e.g., Cowie et al. 1996), while the star-formation timescale from the main sequence $\psi\propto M_\star^{0.8}$ is a slow direct function of the stellar mass or of the time-averaged SFR, namely, $\tau_\star\propto M_\star^{0.2}\propto \langle\psi\rangle^{0.25}$. From a physical point of view, the latter dependence reflects the brevity of the condensation/dynamical timescales within the shallower potential wells of smaller mass halos, that are typically virialized earlier according to the standard structure formation paradigm (see Fan et al. 2010).

The solution of the continuity equation can be worked out under the same
route followed in Aversa et al. (2015), to obtain
\begin{eqnarray}\label{eq|solution}
\nonumber && N(\log M_\star,t) = -\kappa\,(1+\kappa)^{1/\kappa}\, M_\star^{1/\kappa}\,\times\\
\\
\nonumber && \times \int_{0}^t{\rm d}t'~{\partial_{\ln\psi}\over
f_{\tau_\star}}\,\left[{N(\log\psi,t)\, \psi^{-1/\kappa}\over
\tau_\star^{1+1/\kappa}}\right]_{\big{|}\psi_=(1+\kappa)\,M_\star/\tau_\star}~,
\end{eqnarray}
with the shorthand $f_{\tau_\star}\equiv 1+\partial_{\log
\psi}[\log\tau_\star]$; this is in its stand a novel result, although we note that the differences in the outcome relative to a constant or exponential SFR are minor. Similarly, the value of the powerlaw index $\kappa$ is marginally
relevant if varied from the fiducial value $\kappa = 0.5$ within the range
from $0$ (constant SFR) to $1$ (linearly increasing SFR).

\begin{figure}
\epsscale{1}\plotone{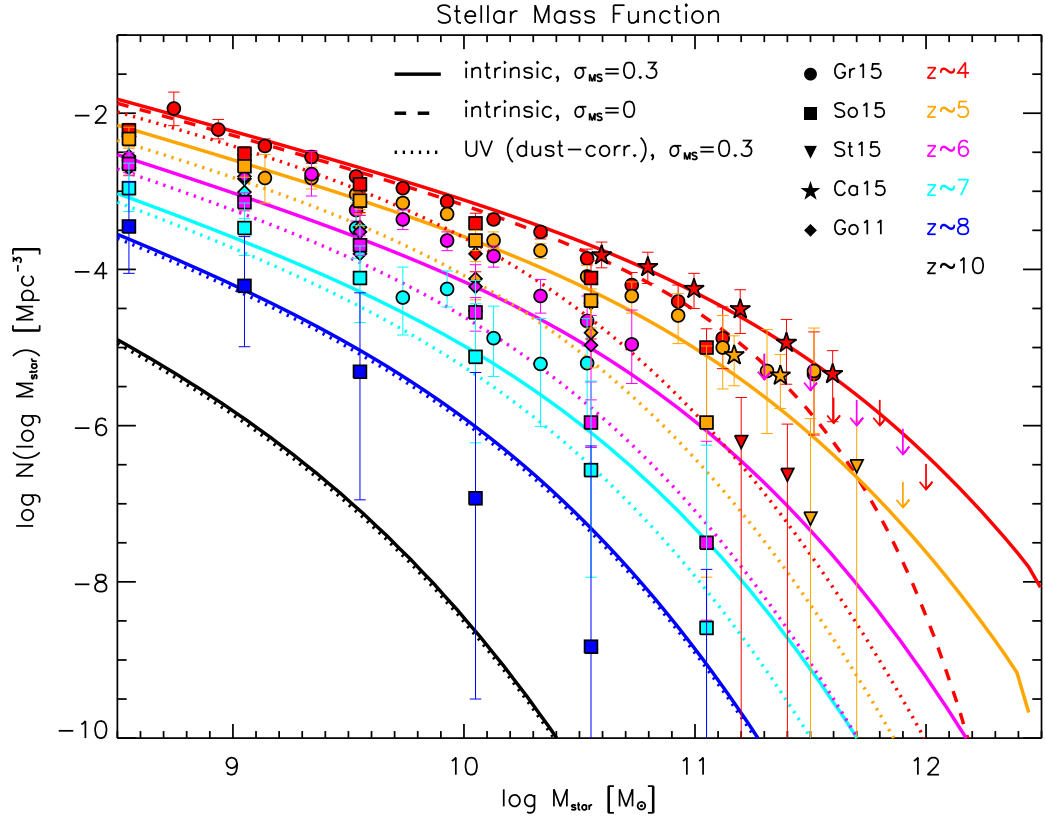}\caption{The stellar mass function
at redshifts $z\approx 4-10$ (color-coded), obtained via the continuity
equation from the intrinsic (solid) or UV-inferred (dotted) SFR function and a scatter of $\sigma_{\rm MS}\approx 0.3$ dex around the median main sequence relationship; at $z\approx 4$ the outcome from the intrinsic SFR function with $\sigma_{\rm MS}\approx 0$ is highlighted by the dashed red line. Data of the stellar mass functions (see text for details) are from Grazian et al. (2015; circles), Song et al. (2015; squares), Stefanon et al. (2015; inverse triangles), Caputi et al. (2015; stars), and Gonzalez et al. (2011; diamonds).}\label{fig|STAR_MF}
\end{figure}

In Fig.~\ref{fig|STAR_MF} we show the resulting stellar mass function at
$z\ga 4$ when using as input our intrinsic SFR functions. The outcome is
compared with the determination of the mass function at $z\approx 4-8$ by
Gonzalez et al. (2011), Grazian et al. (2015), Song et al. (2015), and
Stefanon et al. (2015). The agreement is particularly good with the near-IR selected samples based on HST/WFC3/IR and \textsl{Spitzer} data by Grazian et al. (2015; see also Duncan et al. 2014 and Caputi et al. 2015), when the scatter of $0.3$ dex around the median main sequence relation suggested by Speagle et al. (2014) is taken into account.

Notice that the stellar mass functions by Gonzalez et al. (2011) and Song et al. (2015) are instead obtained from UV-selected samples by combining the observed UV luminosity function with the $M_{\rm UV}-M_\star$ relationship. Even including a scatter of $0.4$ dex in the latter relation as adopted by Song et al. (2015), the stellar mass function is still appreciably lower at the high-mass end with respect to the determination based on near-IR samples; this is due to the underestimation of the luminosity function at the bright end by UV surveys, because of insufficient corrections for dust extinction.

We stress that the number density of massive galaxies $M_\star\approx 10^{11}\, M_\odot$ is still quite high at $z\approx 5$, amounting to about $\approx 10^{-5}$ Mpc$^{-3}$. Then this value is expected to drop around $\approx 5\times 10^{-8}$ Mpc$^{-3}$ at $z\approx 7$, to $\approx 2\times 10^{-9}$ Mpc$^{-3}$ at $z\approx 8$, and to less than $10^{-10}$ Mpc$^{-3}$ at $z\approx 10$; this is mainly due to the rapid falloff of the halo mass function at these high redshifts. However, at $z\ga 6$ data are still uncertain, but reliable measurements will become feasible with next generation instruments; in particular, the \textsl{JWST} will allow to determine stellar masses $M_\star\ga 10^{10}\, M_\odot$ up to $z\sim 7$ (see Caputi 2011). This will eventually allow a detailed validation of the intrinsic SFR function via the continuity equation at these extremely high redshift.

On the other hand, the SFR functions inferred solely from UV data, even corrected for dust extinction, strongly underpredict the observed stellar mass function for $M_\star\ga$ a few $10^{10}\, M_\odot$ at $z\la 5$. This demonstrates that at these redshifts the strong suppression of the bright end in the UV-inferred SFR function with respect to the intrinsic one must be traced back to star formation in dust enshrouded environments, and cannot be related to any form of feedback, like that from SNe or AGNs, that instead would anyhow lower the stellar mass. In the redshift range $z\sim 6-8$ the intrinsic SFR function approaches the UV-inferred one, and particularly so at $z\ga 7$ (cf. Fig.~\ref{fig|SFRlowz}); as a consequence, the continuity equation implies that the stellar mass functions derived from the intrinsic or the UV (dust-corrected) SFR functions are both consistent with the observational determinations within their large uncertainties (cf. Fig.~\ref{fig|STAR_MF}).

\section{Linking to the halo mass via the abundance matching}\label{sec|abmatch}

We now connect the SFR and stellar mass function of active galaxies with the
statistics of the underlying, gravitationally dominant dark matter (DM)
halos. We exploit the abundance matching technique, a standard way of
deriving a monotonic relationship between galaxy and halo properties by
matching the corresponding integrated number densities (e.g., Vale \&
Ostriker 2004, Shankar et al. 2006, Moster et al. 2013; Behroozi et al.
2013).

We derive the relationship $M_\star(M_{\rm H},z)$ between the current stellar mass $M_\star$ and the halo mass $M_{\rm H}$ by solving the equation (see Aversa et al. 2015 for details)
\begin{eqnarray}\label{eq|abmatch}
\nonumber && \int_{\log M_\star}^\infty{\rm d}\log M_\star'\, N(\log M_\star',z) =\int_{-\infty}^{+\infty}{\rm d}\log M_{\rm H}'\times\\
\\
\nonumber && \times  N(\log M_{\rm H}',z)\,{1\over 2}\, {\rm erfc}\left\{{\log[M_{\rm H}(M_\star)/M_{\rm H}']\over \sqrt{2}\,\sigma_{\log M_\star}}\right\}~,
\end{eqnarray}
which holds when a lognormal distribution of $M_\star$ at given $M_{\rm H}$
with dispersion $\sigma_{\log M_\star}$ is adopted; we follow previous
studies based on the abundance matching technique (see references above) and
fiducially take $\sigma_{\log M_\star}\approx 0.15$. In
Eq.~(\ref{eq|abmatch}) the quantity $N(\log M_{\rm H},z)$ is the
galaxy halo mass function, i.e., the mass function of halos hosting one
individual galaxy (see Aversa et al. 2015); actually, for $z\ga 4$ and for
the halo masses of interest here, it coincides with the standard halo mass function from cosmological $N-$body simulations (e.g., Tinker et al. 2008).

The same technique may also be applied to look for a relation $\psi(M_{\rm
H},z)$ specifying the typical SFR $\psi$ in a halo of mass $M_{\rm H}$ at
redshift $z$. However, when dealing with the SFR, one has to take into
account that active galaxies shine with a characteristic star-formation
timescale $\tau_\star(\psi,z)$ which may be smaller than the cosmic time
$t(z)$. In practice, one can still rely on Eq.~(\ref{eq|abmatch}) by
substituting: the current stellar mass with the SFR, i.e.,
$M_\star\rightarrow \psi$; the stellar mass function with the SFR function
divided by the SFR timescale, i.e., $N(\log M_\star,z)\rightarrow
N(\log\psi,z)/\tau_\star(\psi,z)$; and the halo mass function with the halo
creation rate (see Lapi et al. 2013), i.e., $N(\log M_{\rm H},z)\rightarrow
\partial_t^+ N(\log M_{\rm H},z)$.

\begin{figure}
\epsscale{0.49}\plotone{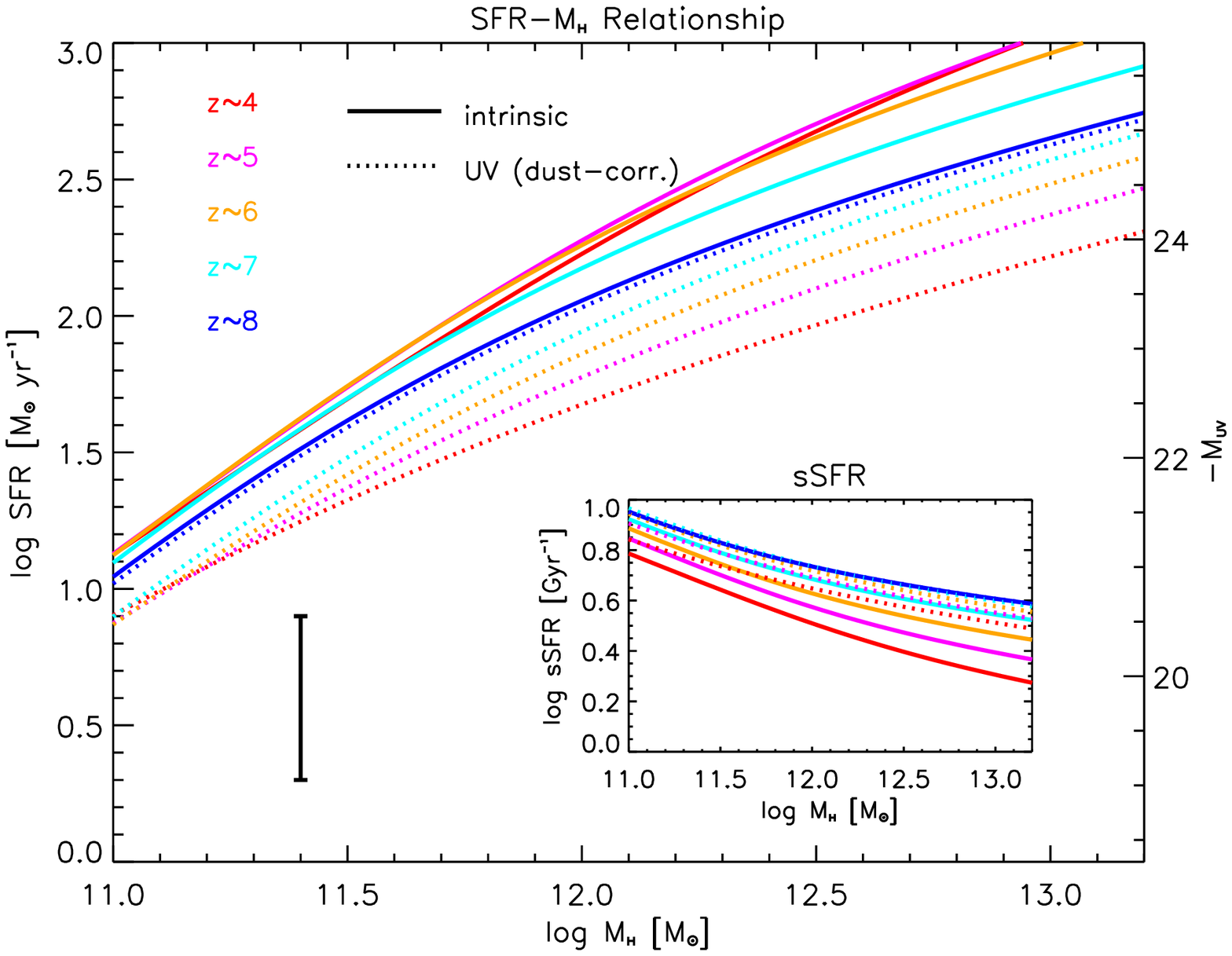}\plotone{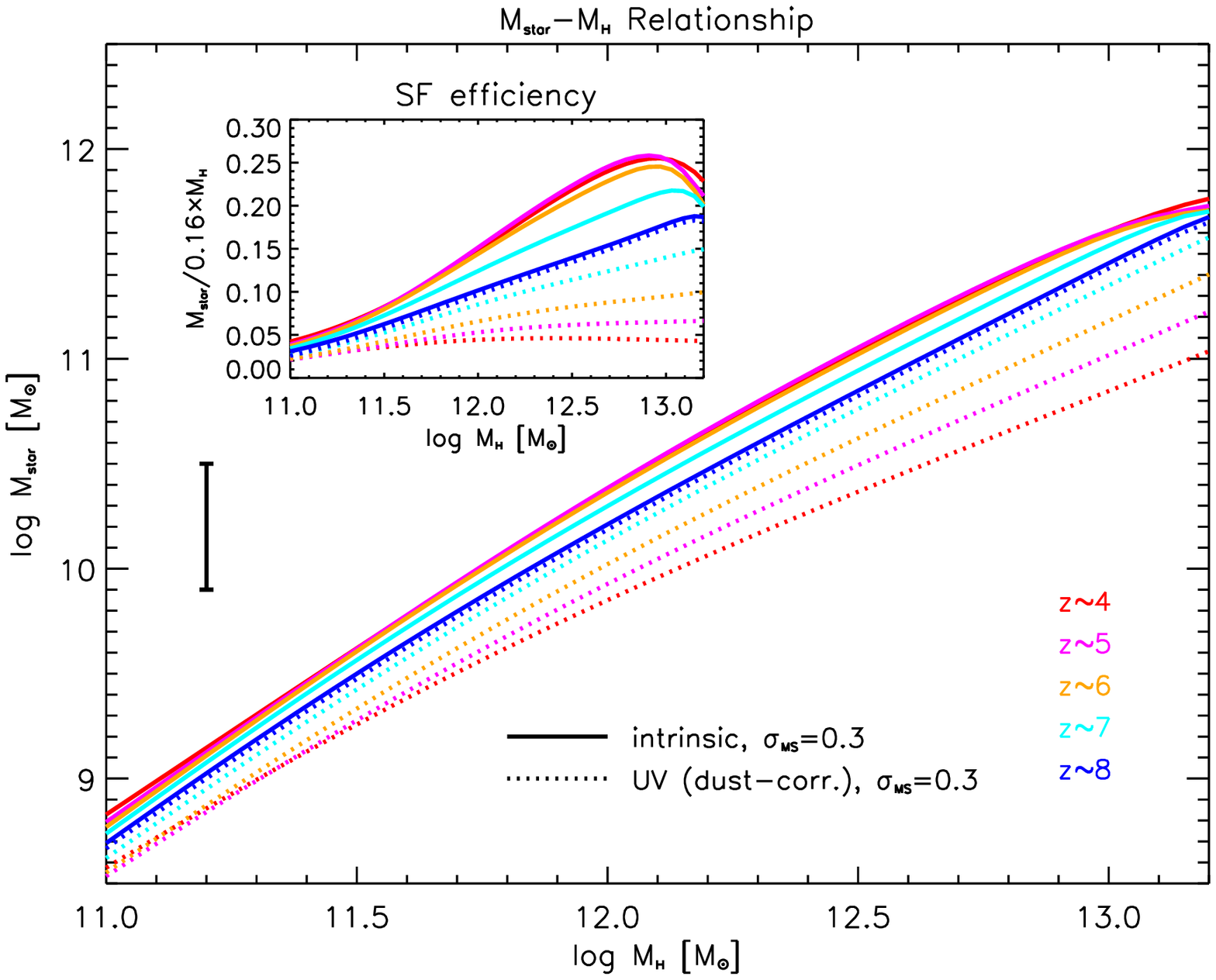}\\\epsscale{0.7}\plotone{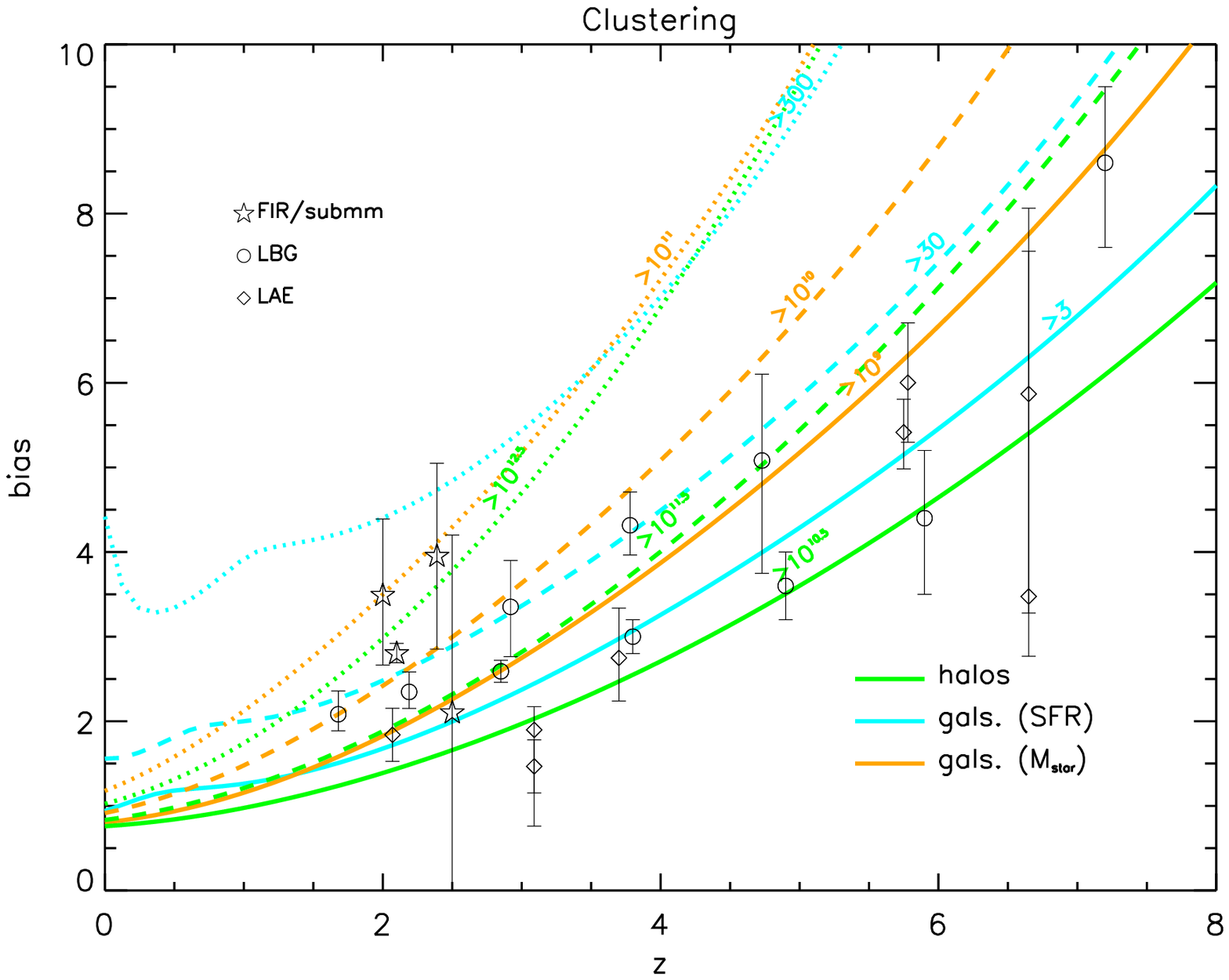}
\caption{Top left panel: the relationship $\psi-M_{\rm H}$ between SFR (right axis shows the corresponding uncorrected UV magnitude) and host halo mass at different redshift $z\approx 4-8$ (color-code), as derived from the abundance matching of the halo mass function to the intrinsic SFR function (solid lines) or to the SFR function inferred from (dust-corrected) UV data (dotted lines); the inset illustrates the corresponding sSFR$=\psi/M_\star$ vs. the halo mass. Top right panel: the same for the relationship $M_\star-M_{\rm H}$ between stellar mass and host halo mass. The inset illustrates the corresponding star-formation efficiency, i.e., the stellar to baryon fraction $M_\star/0.16\times M_{\rm H}$ vs. the halo mass. In both top panels, the error bars represent the typical uncertainty. Bottom panel: the evolution with redshift of the clustering bias; results are shown for halos (green) with DM mass exceeding $10^{10.5}$ (solid), $10^{11.5}$ (dashed) and $10^{12.5}\, M_\odot$ (dotted), for galaxies (cyan) with SFR exceeding $3$ (solid), $30$ (dashed) and $300\, M_\odot$ yr$^{-1}$ (dotted), and for galaxies (orange) with stellar masses exceeding $10^{9}$ (solid), $10^{10}$ (dashed), and $10^{11}\, M_\odot$ (dotted). Data for FIR/(sub-)mm galaxies (stars) are from Weiss et al. (2009), Hickox et al. (2012), Ono et al. (2014), Bianchini et al. (2015), for LBGs (circles) from Ouchi et al. (2004), Adelberger et al. (2005), Lee et al. (2006), Overzier et al. (2006), Bielby et al. (2013), Barone-Nugent et al. (2014), and for LAE (diamonds) from Gawiser et al. (2007), Ouchi et al. (2010), and Guaita et al. (2010).}\label{fig|Abmatch}
\end{figure}

In Fig.~\ref{fig|Abmatch} (top panels) we show the resulting $M_\star-M_{\rm
H}$ and $\psi-M_{\rm H}$ relationships. Note that these relationships refer
to active star-forming galaxies, while Aversa et al. (2015) have presented
the corresponding outcomes for the total population including objects in
passive evolution; as expected, for active galaxies the SFR at given halo mass is higher.

The most remarkable feature of these relationships is the little if no
evolution with redshift at given $M_{\rm H}$; this clearly indicates that the star formation in galaxies at high redshift $z\ga 4$ is regulated by similar, \emph{in-situ} processes (Moster et al. 2013; Aversa et al. 2015), and not by merging or gas infall from cosmological scales. The insets illustrate the
sSFR$=\psi/M_\star$, and the star formation efficiency, i.e., the current
stellar to baryon ratio $M_\star/0.16\times M_{\rm H}$, as a function of
$M_{\rm H}$.

The latter highlights that star formation in galaxies is an extremely
inefficient process, i.e., only a small amount of the available baryon
content of a halo is converted into stars. From a physical point of view,
this is usually interpreted in terms of competition between cooling and
heating processes. In low-mass halos, heating is provided by energy feedback from SN explosions, that regulate star formation at slow rates $\psi\la 10\, M_\odot$ yr$^{-1}$ over timescales of several Gyrs. In massive halos, cooling rates are not significantly offset by SN feedback, yielding the well-known overcooling problem (Cirasuolo et
al 2005; for a recent discussion see Dutton et al. 2015). This
motivated a number of authors (Granato et al. 2004; Di Matteo et
al. 2005; Lapi et al. 2006, 2014) to propose that the star
formation can proceed at much higher levels $\psi\ga 30\, M_\odot$ yr$^{-1}$ over several $10^8$ yr, until the central AGN attains enough power to shine as a quasar, quenching the SFR abruptly and sweeping away most of the gas and dust content (e.g., Shankar et al. 2006; Aversa et al. 2015). On the contrary, neglecting quasar feedback in large halos would produce stellar masses well above the observed values.

We stress that the abundance matching relationships derived on the basis of
the intrinsic and UV-inferred SFR function differ, marginally at $z\ga 7$ but considerably at $z\approx 4$. It is extremely important to take such
differences into account for a proper interpretation of the observational
data in terms of galaxy formation scenarios. For example, consider a galaxy with stellar mass of $M_\star\approx 10^{11}\, M_\odot$ at $z\approx 5$, whose number density is of order $\approx 10^{-5}$ Mpc$^{-3}$ (see Duncan et al. 2014; Grazian et al. 2015). From the $M_\star-M_{\rm H}$ relationship (cf. Fig.~\ref{fig|Abmatch}, top right), the host halo is seen to feature a mass of $M_{\rm H}\approx$ a few $10^{12}\, M_\odot$. Moreover, according to the $\psi-M_{\rm H}$ relationship with its 0.3 dex scatter (cf.
Fig.~\ref{fig|Abmatch}, top left), the SFR turns out to be $\psi\approx
500-1000\, M_\odot$ yr$^{-1}$ when basing on the intrinsic SFR function, but
only of order $\psi_{\rm UV}\approx 50-100\, M_\odot$ yr$^{-1}$ when relying
on the UV-inferred one; the corresponding SFR timescales $M_\star/\psi$
amounts to $\approx 10^8$ yr and $10^9$ yr, respectively. Given that the age
of the Universe at $z\approx 5$ is of order $1.2$ Gyr, the UV-inferred
solution would require star formation to occur well in advance of the initial halo virialization, and/or extreme assumptions on the star formation
efficiency or halo occupation (Steinhardt et al. 2015). On the other hand,
the solution based on the intrinsic SFR function yields a formation redshift
of the host halos $z_{\rm form}\approx 5.4$; the corresponding halo number
density for $M_{\rm H}\approx $ a few $10^{12}\, M_\odot$ reads $\sim
10^{-5}$ Mpc$^{-3}$, in agreement with the stellar mass functions observed at $z\approx 5$ for $M_\star\approx 10^{11}\, M_\odot$ wherefrom the argument
started.

The abundance matching relationships are also fundamental to interpret the
clustering signal associated with high-$z$ dusty galaxies (see
Fig.~\ref{fig|Abmatch}, bottom panel). Specifically, we find that at
$z\approx 4-5$ galaxies endowed with SFR $\psi\ga 100-300\, M_\odot$
yr$^{-1}$ and $M_\star\ga 10^{11}\, M_\odot$ are typically hosted within
halos of $M_{\rm H}\ga$ a few $10^{12}\, M_\odot$, which are extremely biased and clustered. We note that on the basis of the observed clustering signal,
Hildebrandt et al. (2009) and Bian et al. (2013) instead associate the same
halo masses to galaxies selected with UV magnitudes $M_{\rm UV}\la -21$; when using the dust correction based on the UV slope, they estimate a
corresponding SFR $\psi\approx 30-50\, M_\odot$ yr$^{-1}$. This low value of
the intrinsic SFR is underestimated because of an incomplete dust-correction
applied to UV-selected samples (cf. Fig.~\ref{fig|Abmatch}, top left panel),
and would raise again an issue on the star formation timescale (see above;
Steinhardt et al. 2015).

From the top panels of Fig.~\ref{fig|Abmatch} it is apparent that, when the abundance matching is performed by exploiting the intrinsic SFR function, the evolution with redshift of SFR and stellar mass at fixed halo mass is small and well within the errors determined by observations. On the other hand, the evolution is amplified when the abundance matching is performed by exploiting  the (dust-corrected) UV-inferred SFR function. The latter case would imply an increasing star-formation efficiency with redshift, which is reminiscent of the claim by Finkelstein et al. (2015b). On the contrary, we find no evolution (within errors) of the star-formation efficiency when basing on the intrinsic SFR function.

\section{Hunting high-$z$ dusty starforming galaxies}\label{sec|hunting}

In the above we have stressed the relevance of probing the statistics of galaxies at $z\sim 4-6$ with SFRs $\psi \ga 10^2\, M_\odot$ yr$^{-1}$, that contribute substantially to the high-mass end of the stellar mass function. Since most of these galaxies with high SFRs are likely dust-enshrouded, exploiting IR data is mandatory to fully assess the intrinsic SFR function. How to achieve this goal in practice, given the current and upcoming observational facilities in the far-IR/(sub-)mm and/or UV band, constitutes the issue addressed below.

\subsection{Selecting dusty galaxies in the far-IR/(sub-)mm band}\label{sec|FIRnature}

\begin{figure}
\epsscale{0.7}\plotone{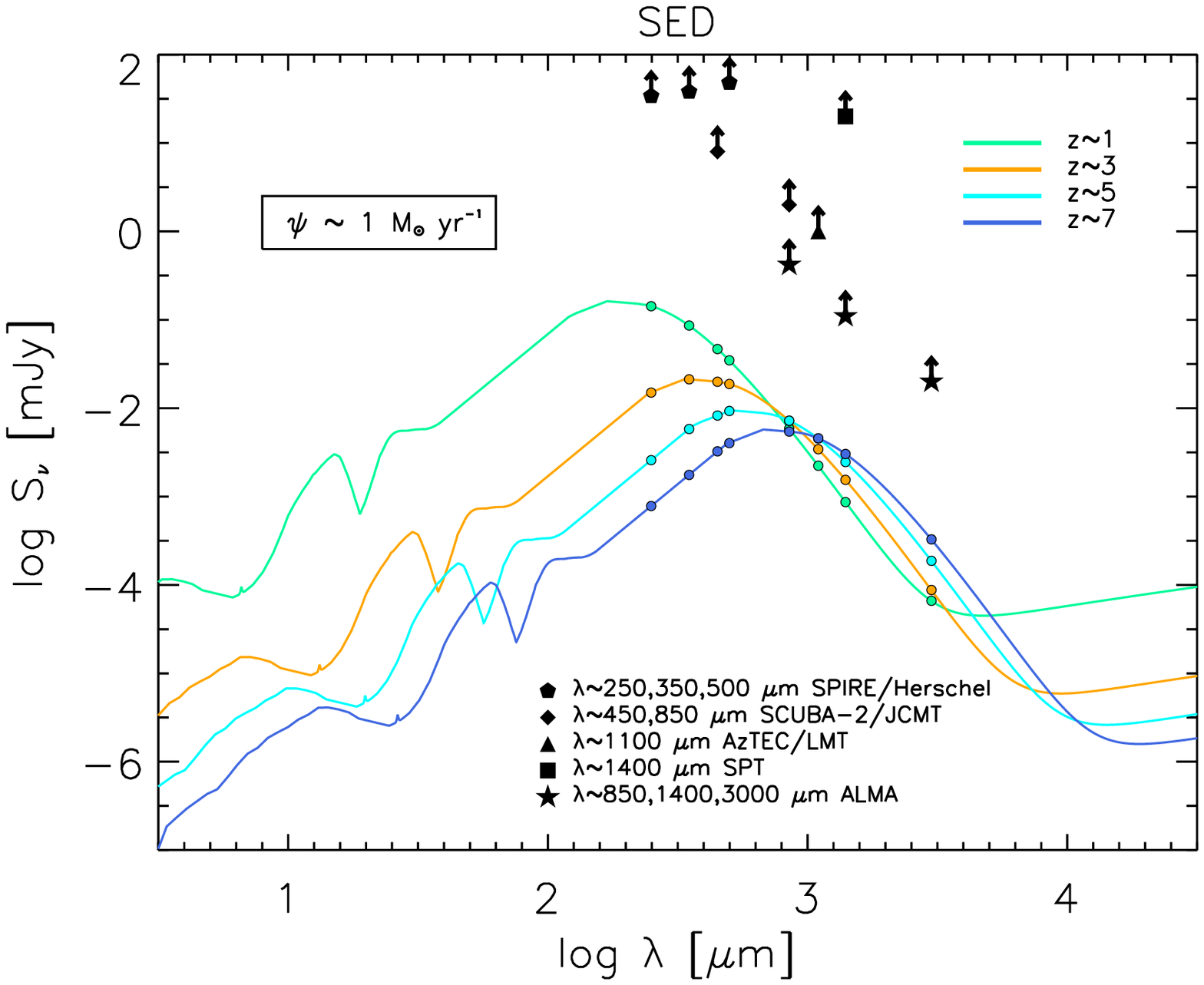}\\\plotone{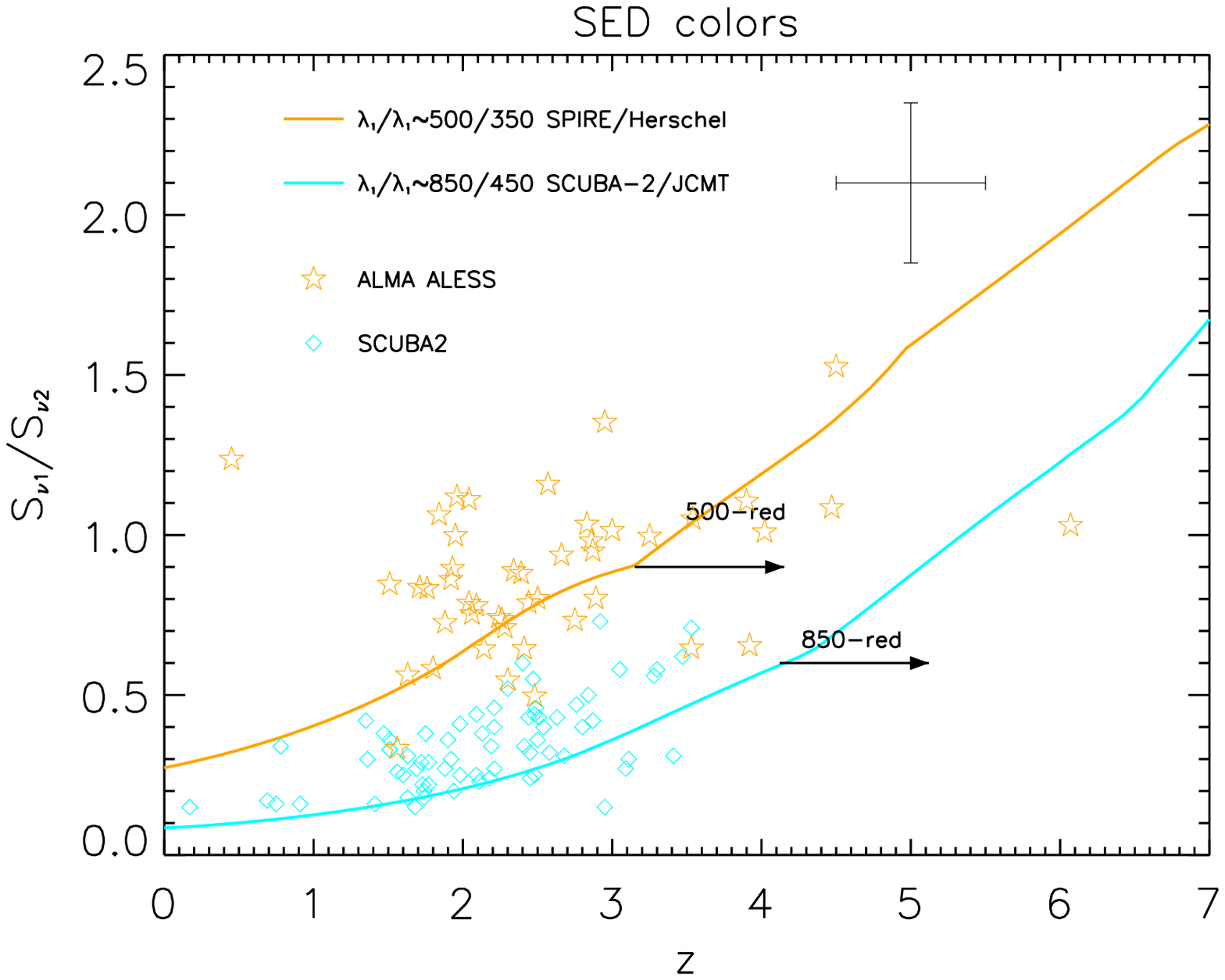} \caption{Top panel:
the SED of a typical high$-z$, dust-obscured star-forming galaxy located at
redshift $z\approx 1$ (green), $3$ (orange), $5$ (cyan), and $7$ (blue),
normalized to a SFR $\psi=1\, M_\odot$ yr$^{-1}$ in the range $\lambda\approx 3-1100\, \mu$m. Colored symbols illustrate the values of the SED at different operating wavelengths for various instruments: SPIRE/\textsl{Herschel} (circles), \textsl{SCUBA-2} (diamonds), \textsl{AzTEC} (triangles), \textsl{SPT} (squares), and \textsl{ALMA} (stars). The corresponding $5\sigma$ sensitivities are shown by the black symbols with arrows. Bottom panel: SED colors $S_{\nu,1}/S_{\nu,2}$ as a function of redshift, exploited for \textsl{Herschel} and \textsl{SCUBA-2} preselection of dusty galaxies; different lines refer to $500/350$ (orange) and $850/450$ (cyan) colors. Arrows indicate the redshift ranges where red galaxies are preferentially located. Data are from \textsl{ALMA}/ALESS by Swinbank et al. (2014; orange stars) and from \textsl{SCUBA-2} by Koprowski et al. (2015; cyan diamonds); in the upper right corner the typical data uncertainty on the median is reported.}\label{fig|sed}
\end{figure}

As a starting point, in Fig.~\ref{fig|sed} (top panel) we illustrate the redshift evolution for our reference SED (see Sect.~\ref{sec|counts}). The SED has been normalized so that the far-IR emission in the range $3-1100\, \mu$m corresponds to a SFR of $\psi\approx 1\, M_\odot$ yr$^{-1}$.

We have illustrated the positions on the SED of the observational wavelengths for various instruments of interest here : $250$, $350$, and $500\, \mu$m for the SPIRE instrument on board of \textsl{Herschel}; $450$ and $850\, \mu$m for the \textsl{SCUBA-2} instrument at the JCMT; $\sim 1100\, \mu$m for the \textsl{AzTEC} at the LMT; $1400\,\mu$m for the \textsl{SPT}; and $850$, $1400$ and $3000\, \mu$m for \textsl{ALMA}. We have also highlighted $5\sigma$ detection limits for such instruments (attained in the deepest large-scale surveys undertaken so far or upcoming): $S_{250}\approx 35$ mJy, $S_{350}\approx 40$ mJy, and $S_{500}\approx 50$ mJy for SPIRE; $S_{450}\approx 8$ mJy and $S_{850}\approx 2$ mJy for \textsl{SCUBA-2}; $S_{1100}\approx 1$ mJy for \textsl{AzTEC}; $S_{1400}\approx 20$ mJy for \textsl{SPT}; and $S_{850}\approx 0.42$, $S_{1400}\approx 0.11$, and $S_{3000}\approx 0.02$ mJy for \textsl{ALMA} (500 hours on $100$ arcmin$^2$).

The accurate determination of the spectroscopic redshift for a large sample
of dusty galaxies is a major problem. In order to probe the bright end of the SFR function at $z\ga 3$, a strategy could be to preselect high-redshift
sources using flux/color criteria from surveys conducted with
\textsl{Herschel} or \textsl{SCUBA-2}, and then perform a
more accurate photometric (or even spectroscopic) redshift determination with observations on source by, e.g., \textsl{AzTEC} and \textsl{ALMA}.

Commonly used preselection criteria (e.g., Dowell et al. 2014; Asboth et al. 2016) for high-redshift sources based on \textsl{Herschel} photometry involve to look for `$350-$peakers' defined as sources with $S_{350}/S_{250}\ga 1$ and $S_{500}/S_{350}\la 1$, or `$500-$risers' defined as sources with $S_{350}/S_{250}\ga 1$ and $S_{500}/S_{350}\ga 1$. Actually, the uncertainties in the flux measurements make the distinction between peakers and risers quite loose; moreover, at $z\ga 4$ the channel at $250\, \mu$m refers to restframe wavelengths $\lambda\la 50\, \mu$m, where details of the SED due to different dust properties and a possible contribution from an AGN component can be relevant. Thus here we mainly focus on the color $S_{500}/S_{350}$. In Fig.~\ref{fig|sed} (bottom panel) we plot its evolution with redshift, and compare it with the measurements from \textsl{ALMA}/ALESS by Swinbank et al. (2014) finding a reasonable agreement within the large uncertainties. Sources with $S_{500}/S_{350}\ga 0.8$ are high-redshift $z\ga 3$ candidates.

However, the rather high limiting fluxes of \textsl{Herschel} cannot probe the SFR function much above $z\sim 5$, since even sources with SFR $\psi\ga 10^3\, M_\odot$ yr$^{-1}$ are too faint to be detected. To go much beyond $z\ga 4$, the preselection based on the color $S_{850}/S_{450}$ from \textsl{SCUBA-2} photometry is much more efficient. In Fig.~\ref{fig|sed} (bottom panel) we plot its evolution with redshift, and compare it with the measurements from \textsl{SCUBA-2} by Koprowski et al. (2015) finding again a reasonable agreement. It is seen that the color condition $S_{850}/S_{450}\ga 0.6$ can be exploited to preselect candidate galaxies at $z\ga 4$.

\begin{figure}
\epsscale{0.7}\plotone{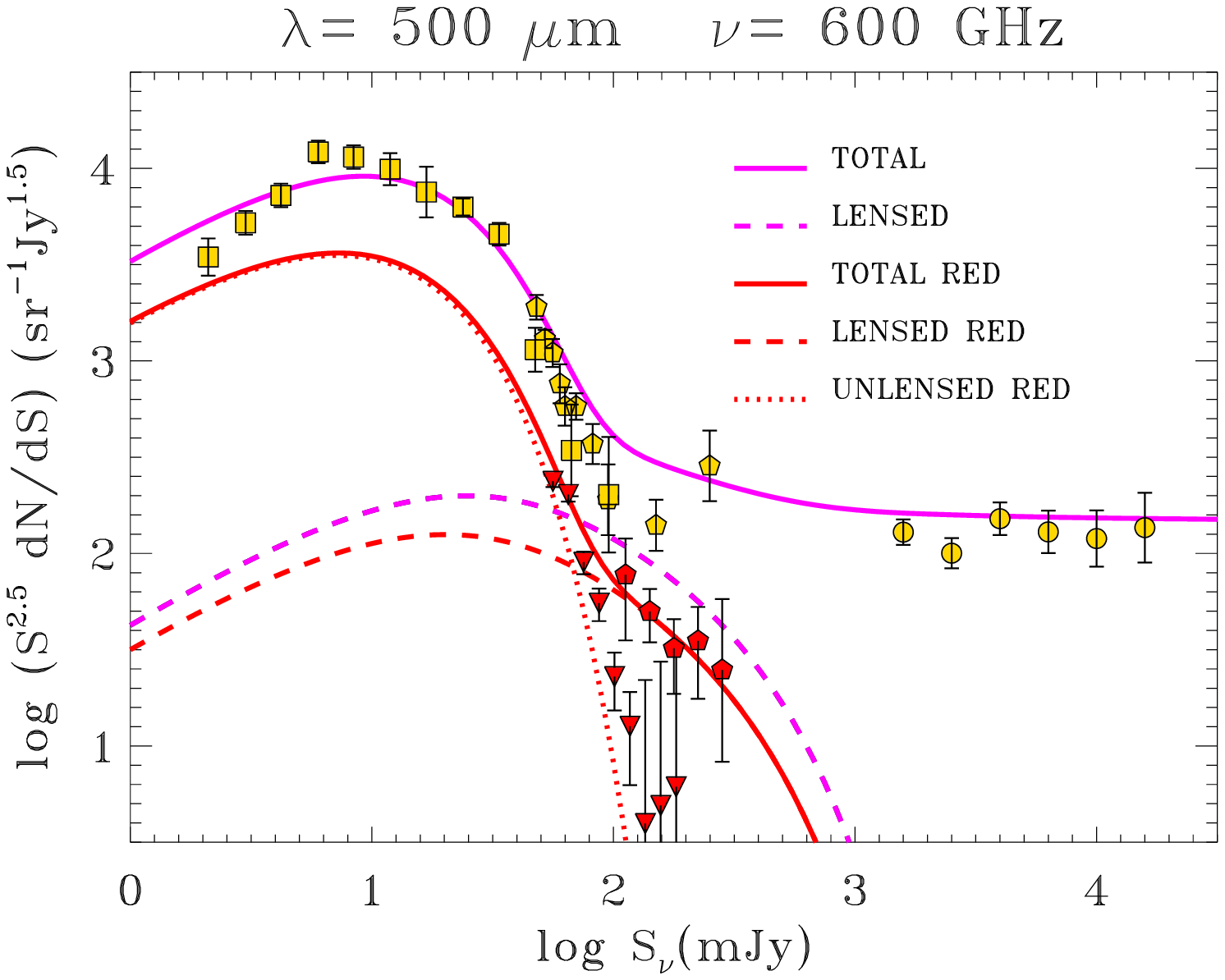}\\\plotone{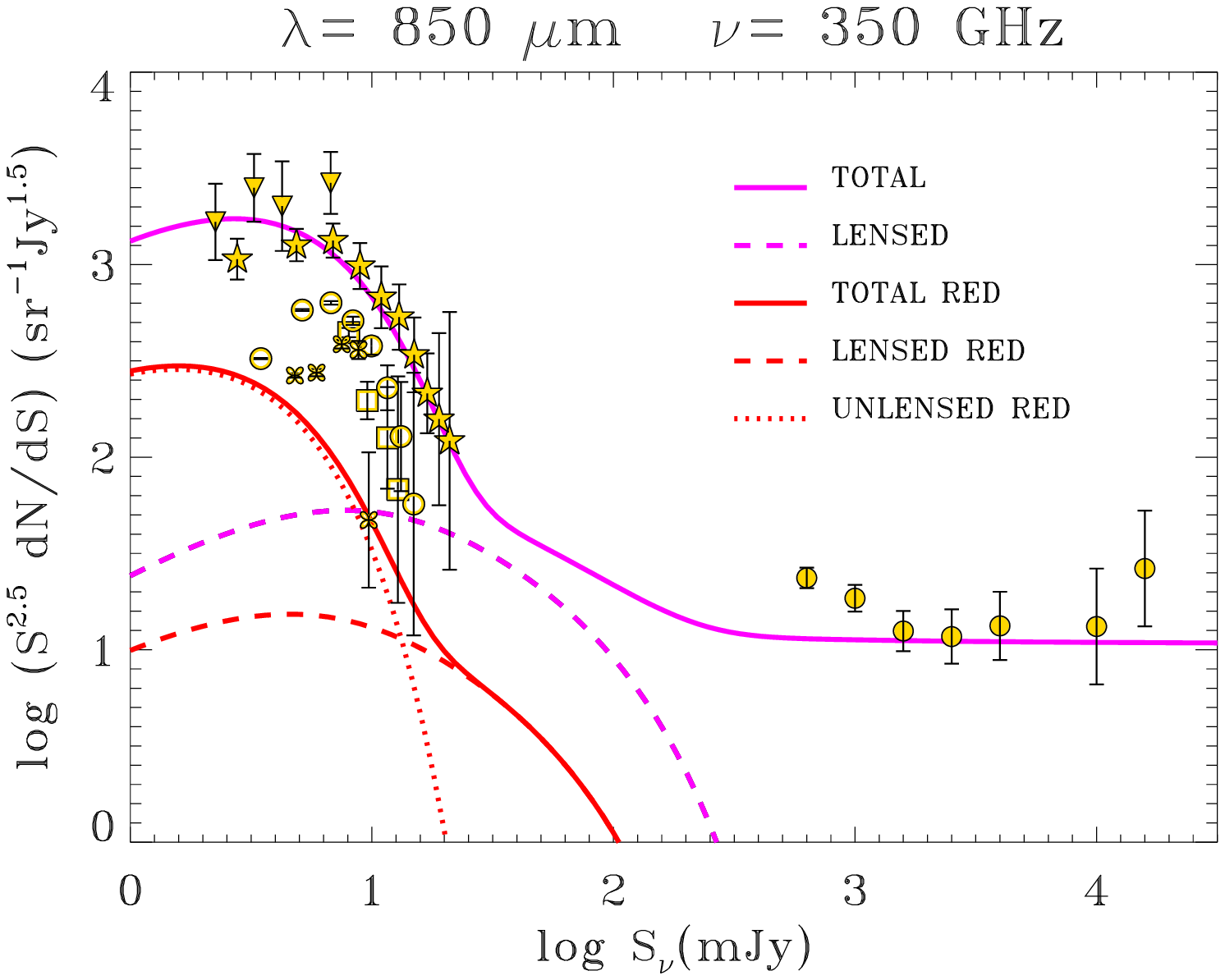}
\caption{Euclidean-normalized differential number counts at $500$ (top) and $850\, \mu$m (bottom). Magenta lines refer to the counts derived from our intrinsic SFR function; the contribution to the total counts (solid) from strongly lensed galaxies (dashed) is highlighted. The counts of red galaxies are illustrated in red, with the solid lines referring to the total, the dotted lines to unlensed counts and the dashed lines to the gravitationally lensed sources. Data (gold symbols) are as in Fig.~\ref{fig|diffcounts}. At $500\, \mu$m data for red galaxies are from \textsl{Herschel}/HerMES by Asboth et al. (2016, red inverse triangles; see also Dowell et al. 2014) and of candidate high-$z$ lenses from \textsl{Herschel}/ATLAS by Negrello et al. (2016, red pentagons; see also Wardlow et al. 2013, Nayyeri et al. 2016).}\label{fig|intcounts}
\end{figure}

In Fig.~\ref{fig|intcounts} we show our predictions for the differential counts at $500$ and $850\, \mu$m for the red sources preselected according to the color criteria discussed above. At $500\, \mu$m the counts of unlensed red sources with $S_{500}\la 100$ mJy agree with the determination by Asboth et al. (2016; see also Dowell et al. 2014) while those of lensed red sources well compare with the candidate lenses in the \textsl{Herschel}/ATLAS survey selected via their red colors by Negrello et al. (2016, in preparation; see also Nayyeri et al. 2016 and Wardlow et al. 2013 for analogous studies in the \textsl{Herschel}-HeLMS+HerS and \textsl{Herschel}-HerMES surveys).

\begin{figure}
\epsscale{0.7}\plotone{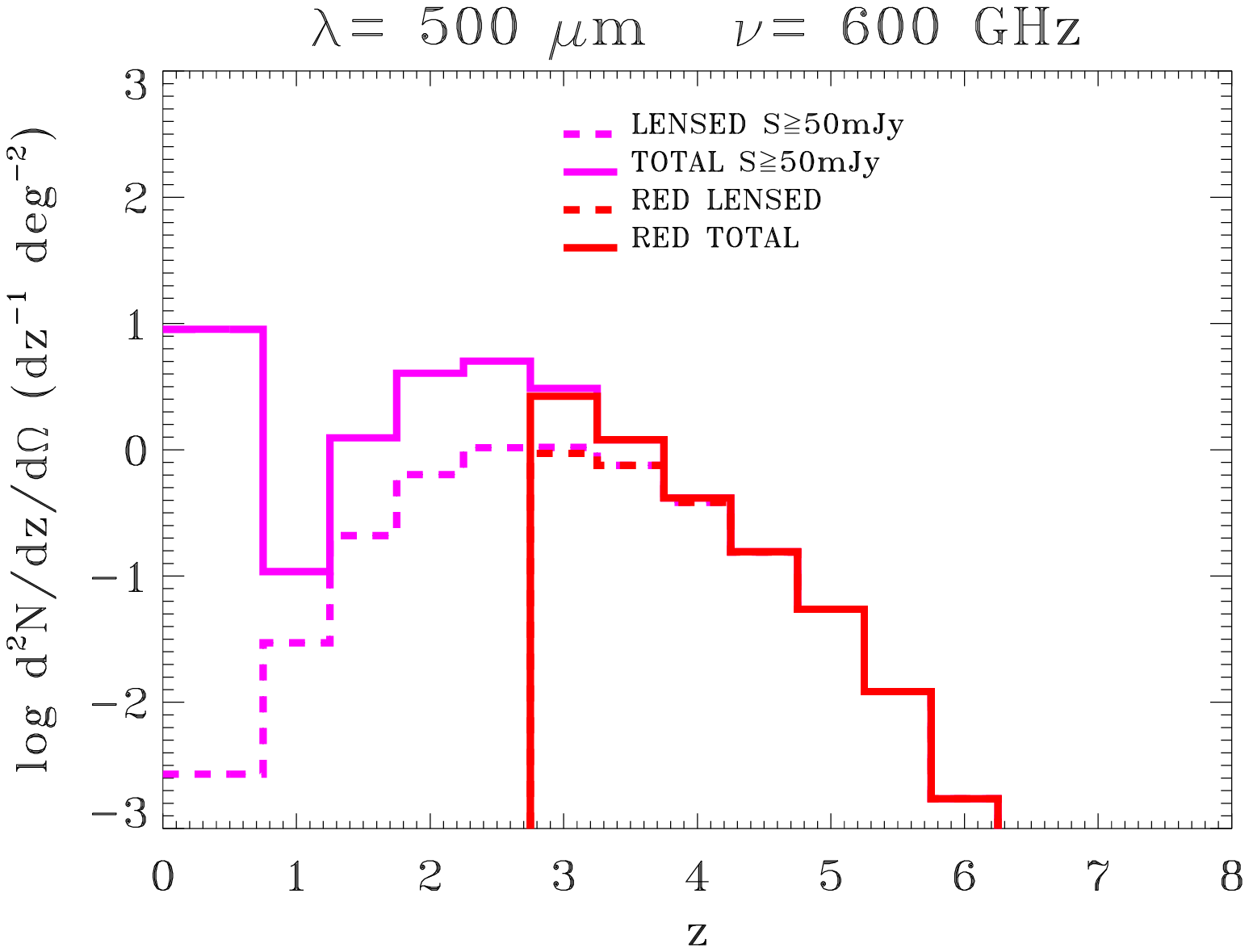}\\\plotone{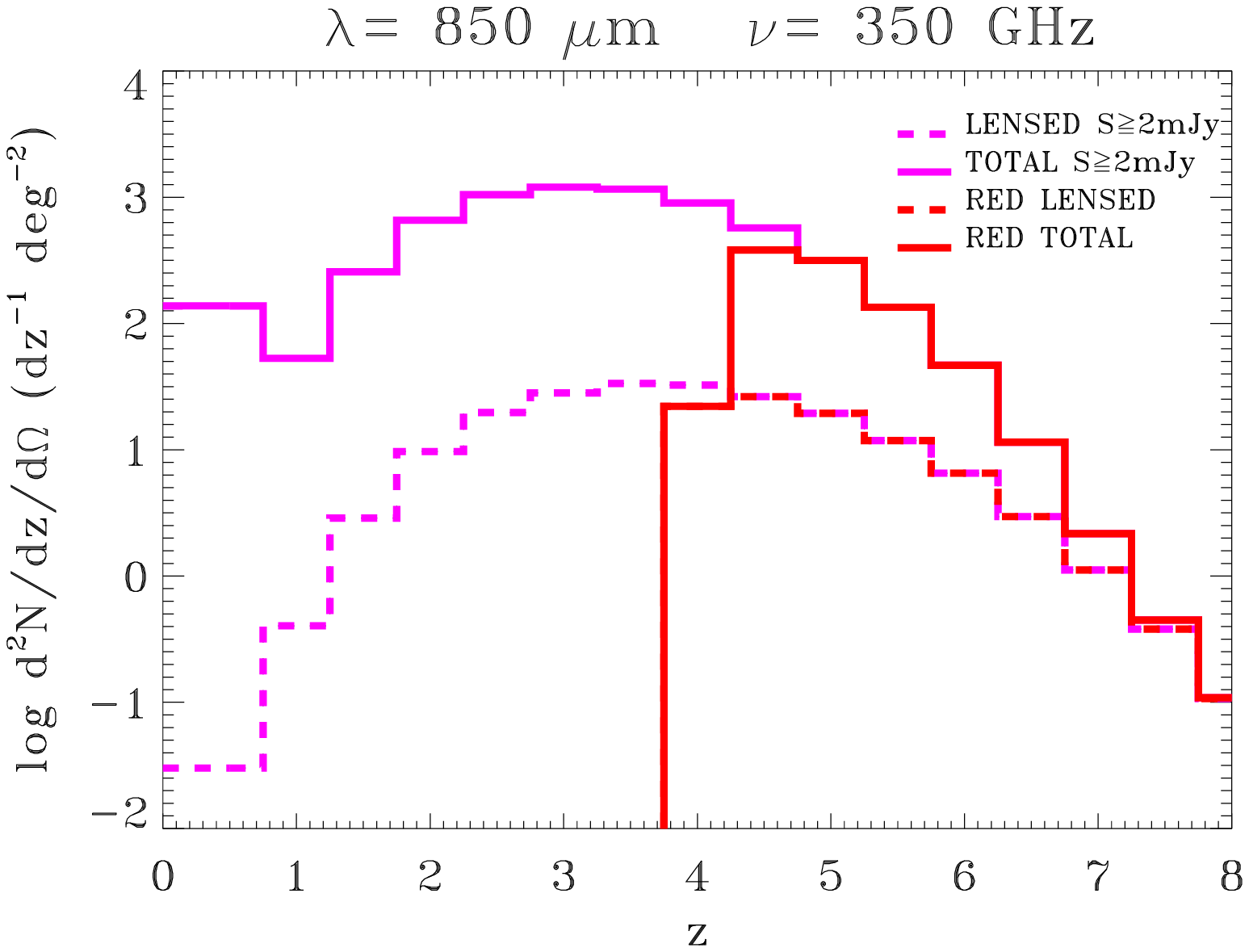}
\caption{Redshift distributions at $500$ (top panel) and $850\,\mu$m (bottom panel). At $500\, \mu$m magenta lines refer to the \textsl{Herschel} limiting flux of $\approx 50$ mJy, with the contribution to the total (solid) from strong galaxy-scale gravitational lensing (dashed) highlighted; the same for red sources is shown in red. At $850\, \mu$m the magenta lines refer to a limiting flux of $2$ mJy (again red lines refer to red sources).}\label{fig|zdist}
\end{figure}

In Fig.~\ref{fig|zdist} (top panels) we present the corresponding redshift distributions. At $500\,\mu$m (top left) red high-$z$ candidates are mostly located at redshift $z\ga 3$, featuring SFRs $\psi\ga 300\, M_\odot$ yr$^{-1}$. Interestingly, strong gravitational lensing by foreground galaxies broadens the tail of their redshift distribution toward $z\approx 5-6$; we stress that most of the lensed sources are amplified by relatively modest factors $\langle\mu\rangle\approx 2-5$, so that they would on average feature SFRs still of $\psi\ga 100\, M_\odot$ yr$^{-1}$. At $850\, \mu$m (top right) the color selection based on $850$ to $450$ flux ratio picks up objects at $z\ga 4$, with a tail extending out to $z\approx 7-8$. Since the \textsl{Herschel} surveys at $350$ and $500\, \mu$m cover an overall area of $\sim 1000$ deg$^{-2}$, their data can be mined to pick out $\ga 1000$ gravitationally lensed galaxies (see Gonzalez-Nuevo et al. 2012), which, in addition to their cosmological interest (e.g., Eales 2015), can be also used to estimate the SFR distribution function up to $z\sim 6$.

Once the high$-z$ candidates have been preselected, the photometric data from \textsl{Herschel} or \textsl{SCUBA-2} have then to be supplemented with observations at longer wavelengths by, e.g., \textsl{AzTEC} and \textsl{ALMA}; the latter instrument can be also exploited to go for a spectroscopic redshift determination, but this plainly requires more observing time, and preferentially a precise estimate of the photometric redshift to choose the most-suited observational band.

\subsection{Dusty galaxies are not lost in the UV band}\label{sec|UVnature}

In the above we have demonstrated that the intrinsic SFR functions as
validated via the (sub-)mm counts and the continuity equation are largely
underestimated by UV data, especially at the bright end for $z\la 7$.
However, next we show that such dusty galaxies can be efficiently probed by combining current UV surveys with upcoming far-IR/(sub-)mm and radio observations.

\begin{figure}
\epsscale{.7}\plotone{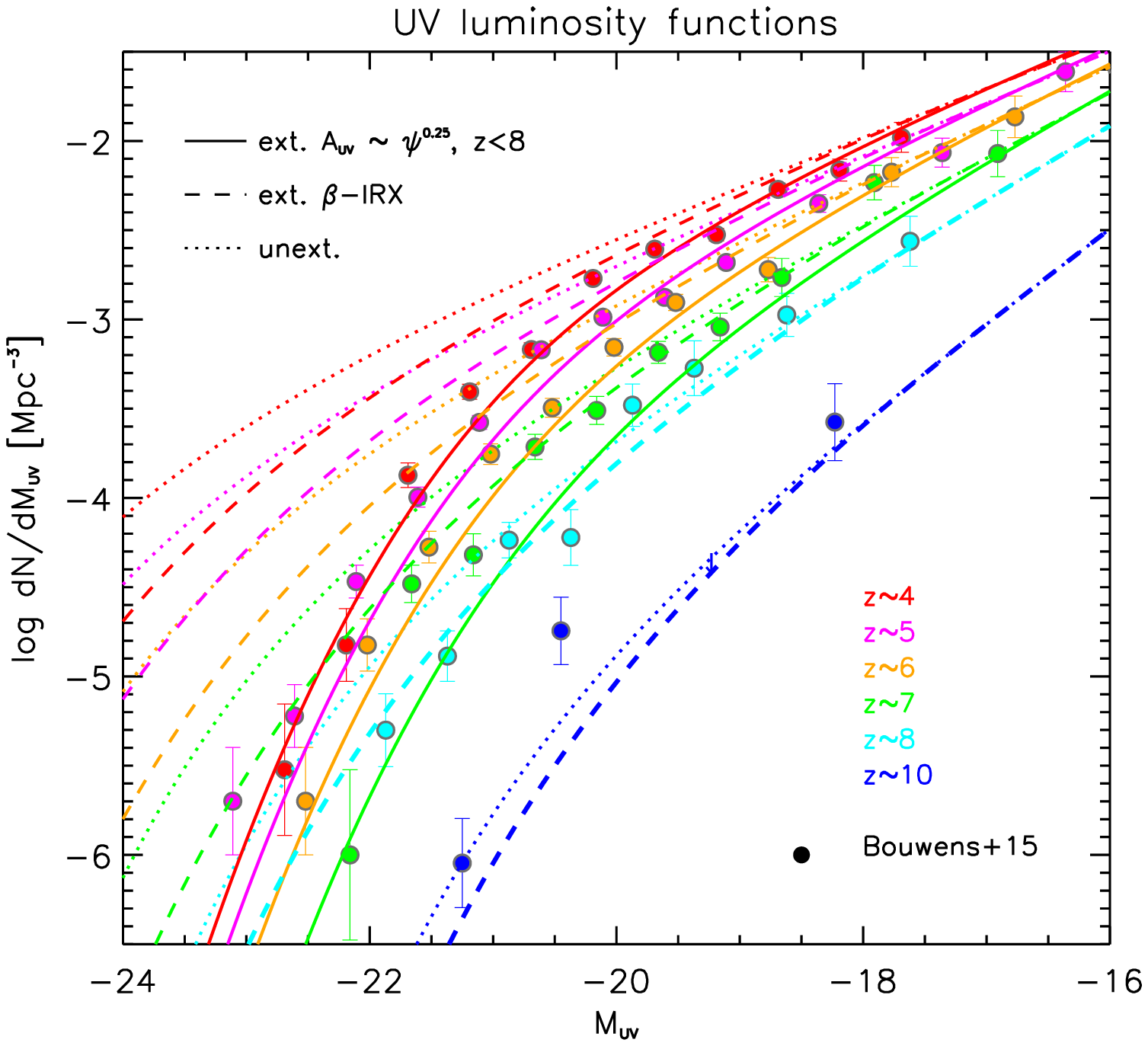}\\\plotone{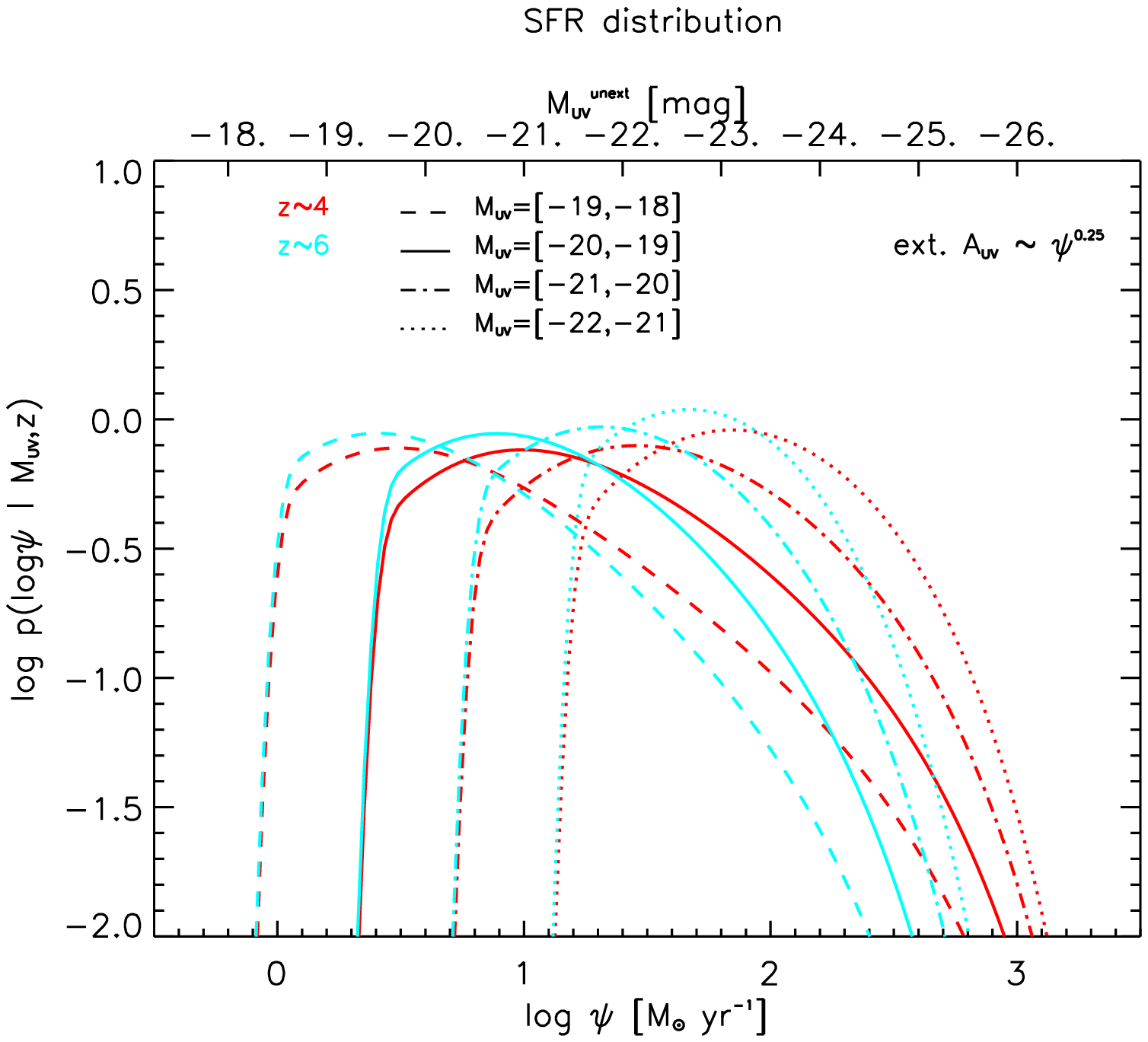}\caption{Top panel: UV luminosity function at different redshift $z\approx 4-10$ (color-coded),
as reconstructed from our intrinsic SFR function by not correcting for dust
extinction (dotted lines), correcting via the standard $\beta_{\rm UV}$-IRX
relation (dashed lines), and via the simple relationship $A_{\rm
UV}=\psi^{0.25}$ (solid lines) with a scatter of $1$ mag at given $\psi$ (for $z<8$). Data points (circles) are from Bouwens et al. (2015). Bottom panel:
normalized SFR distribution of galaxies in the observed UV magnitude
bins centered at $M_{\rm UV}\approx -18.5$ (dashed lines), $-19.5$ (solid), $-20.5$ (dot-dashed) and $-21.5$ (dotted) at redshifts $z\sim 4$ (red) and $6$ (cyan). The upper axis refer to the unextincted UV magnitude $M_{\rm
UV}^{\rm unext}$ associated to the intrinsic SFR $\psi$. The extinction law $A_{\rm UV}=\psi^{0.25}$ with a scatter of $1$ mag at given $\psi$ has
been adopted.}\label{fig|UVnature}
\end{figure}

As a starting point, in Fig.~\ref{fig|UVnature} (top panel) we present the UV luminosity functions at $z\ga 4$, as reconstructed from the intrinsic SFR function by using various prescriptions for dust extinction. We start by showing that the outcome when no correction is applied considerably overestimate the UV luminosity function for any redshift $z\la 7$ at the bright end $M_{\rm UV}\la -19$. This occurs even when the standard correction based on the $\beta_{\rm UV}$-IRX relation is adopted. This is because, as shown by several authors (e.g., Reddy et al. 2012, Davies et al. 2013; Fan et al. 2014; Coppin et al. 2015) thanks to mid-/far-IR observations of UV-selected galaxies at redshift $z\sim 2-4$, the attenuation values of galaxies with observed UV magnitude $M_{\rm UV}\la -21$ are strongly in excess with respect to those estimated basing on the $\beta_{\rm UV}$-IRX relation, and feature a very large dispersion. The standard interpretation is that star formation occurs preferentially within heavily dust-enshrouded molecular clouds, while the UV slope mainly reflects the emission from stars obscured by the diffuse, cirrus dust component (see Silva et al. 1998; Coppin et al. 2015; Reddy et al. 2015). We notice that on approaching $z\approx 8$ the $\beta-$IRX corrected luminosity function converges toward the unextincted one, at least down to $M_{\rm UV}\ga -21.5$, corresponding to an intrinsic SFR $\psi\la 30\, M_\odot$ yr$^{-1}$; this suggests that for these galaxies the timescale required to accumulate substantial amount of dust becomes longer than their age.

All in all, for $z\la 7$ and $M_{\rm UV}\la -19$, an attenuation larger than that derived on the basis of the $\beta-$IRX relation is needed to recover the UV luminosity function from the intrinsic SFR function.

Observationally, at $z\sim 2$ the the UV attenuation $A_{\rm UV}$ is found to directly correlate with the SFR, though with a large dispersion of about $1$ mag (or $0.4$ dex in $\log$ IRX; e.g., Reddy et al. 2010); the attenuations are already significant with values $A_{\rm UV}\approx 1.5-2.5$ mag (or IRX values of $\approx 4-7$) for SFRs $\psi\approx 30-50\, M_\odot$ yr$^{-1}$. A heuristic rendition of such an observed correlation for $\psi\ga 30\, M_\odot$ yr$^{-1}$ reads
\begin{equation}\label{eq|UVattenuation}
A_{\rm UV} = \psi^{0.25}~;
\end{equation}
Note that, via stacking analysis of $850\, \mu$m emission from Lyman Break Galaxies in the \textsl{SCUBA-2} Cosmology Legacy Survey, Coppin et al. (2015; cf. their Table 2) finds at $z\sim 3-5$ average values of SFR $\psi\approx 70-130\, M_\odot$ yr$^{-1}$ and UV attenuations $A_{\rm UV}\approx 2.3-2.5$, broadly consistent with Eq.~(\ref{eq|UVattenuation}).
Note that the relation expressed by  Eq.~(\ref{eq|UVattenuation}) represents an average UV attenuation defined as $A_{\rm UV}\equiv 2.5\, \log(1+L_{\rm IR}/L_{\rm UV})$ in terms of the integrated UV and IR luminosities, and as such it is a basic quantity whose estimate does not require a full radiative transfer approach (e.g., Meurer et al. 1999; Reddy et al. 2015).

In Fig.~\ref{fig|UVnature} (top panel) we show this to map remarkably well the intrinsic SFR functions onto the observed UV luminosity function over the redshift range $z\approx 4-7$. The outcome on the luminosity function is mostly sensitive to the scatter of the above relation; the behavior at the bright end of the luminosity function constrains it to be within $\pm 0.2$ of the best-fit value of $1$ mag. We caveat that the above equation does not include a dependence on metallicity $Z$, which may well affect the dust abundance. As a matter of fact, Reddy et al. (2010) find a direct dependence between $A_{\rm UV}$ and the metallicity $Z$, with the former becoming appreciable when $Z\ga Z_\odot/3$. A combined dependence $A_{\rm UV}\propto \psi^{\alpha}\, Z^{\beta}$ including both the SFR and metallicity has been considered by Mao et al. (2007) and Cai et al. (2014) basing on the $M_{\rm UV}$ vs. $E(B-V)$ relation by Shapley et al. (2001) and the UV luminosity functions at different redshift. Actually in their approach $\psi(\tau)$, $Z(\tau)$ and hence $A_{\rm UV}(\tau)$ are function of the galactic age $\tau$, but for galaxies with quite robust SFR $\psi\ga 30-50\, M_\odot$ yr$^{-1}$ the SFR is roughly constant and the metallicity saturates rapidly for $\tau\ga$ a few $10^7$ yr to slightly subsolar values; all in all, their time-averaged relation is very close to Eq.~(\ref{eq|UVattenuation}).

In Fig.~\ref{fig|UVnature} (bottom panel), we represent the SFR distribution
(areas under curves are normalized to $1$) for galaxies selected with a given observed UV magnitude at redshift $z\sim 4-6$; the upper axis refers to the
unextincted UV magnitude corresponding to a given SFR $\psi$. Plainly, the
rapid truncation of the distributions to the left of the peak occurs because
values of the SFRs yielding an unexctincted UV magnitude fainter than the
observed are not allowed ($A_{\rm UV}>0$ must hold). The decrease of the
distributions to the right of the peak reflects the convolution between the
intrinsic SFR function and the adopted attenuation law with its large dispersion. The distributions tend to be narrower at higher redshift and for brighter observed UV magnitudes, due to the evolution of the intrinsic SFR function at the high-SFR end, and to the decrease of the extinction with increasing redshift.

Basing on the star formation main sequence, we expect that most of the dusty galaxies with intrinsic SFR $\psi\ga 100\, M_\odot$ yr$^{-1}$ that appear as faint UV objects with $M_{\rm UV}\ga -21$ also feature large stellar masses $M_\star\ga$ a few $10^{10}\, M_\odot$. As such, they appear as upper outliers in the $M_\star-M_{\rm UV}$ diagram (see Duncan et al. 2014, Grazian et al. 2015; Song et al. 2015; Coppin et al. 2015). Note that a similar location could also be occupied by almost passively evolving galaxies, but their number at $z\ga 4$ is expected to be small, since the star formation timescales implied by the main sequence are close to the age of the Universe. As a consequence, a significant fraction of the outliers are expected to be highly star-forming, massive galaxies, and as such constitute particularly suited targets for far-IR and (sub-)mm observations.
At redshift $z\ga 6$ the same regime will be explored by \textsl{JWST}.

The broad shape of the SFR distributions implies that dusty, strongly star-forming galaxies with $\psi\ga 30\, M_\odot$ yr$^{-1}$ are not lost in the UV, but rather are moved by their strong attenuation $A_{\rm UV}\ga 2.3$ at fainter magnitudes; although they are outnumbered by the intrinsically faint and poorly attenuated galaxies, nevertheless they can be singled out following the strategy proposed below. In Table~2 we present the number per sq. arcmin of dusty, UV-selected galaxies expected per observed magnitude bin, for a given threshold in SFR. The numbers decrease quite rapidly with increasing redshift and increasing SFR threshold. For example, considering that the current areas surveyed in the UV amount to $\approx 10^3$ arcmin$^2$ (see Bouwens et al. 2015; their Table 1), the expected numbers are around several hundreds of galaxies with SFR $\psi\ga 100\, M_\odot$ yr$^{-1}$ at $M_{\rm UV}\la -17$ and $z\approx 4$; this number decreases to several tens at $z\approx 6$ for $\psi\ga 100\, M_\odot$ yr$^{-1}$, and to a hundred at $z\approx 4$ for $\psi\ga 300\, M_\odot$ yr$^{-1}$.

These UV data could be exploited in combination with (sub-)mm and/or radio observations to reconstruct the bright end of the intrinsic SFR function (e.g., Barger et al. 2014). The strategy involves to observe the areas $\la 10^3$ arcmin$^2$ of current UV surveys (see Bouwens et al. 2015) in the (sub-)mm and/or radio band. On one hand, this will allow to measure  unbiasedly the intrinsic SFR of strongly dust-obscured galaxies from (sub-)mm/radio data; on the other hand, the cross matching with the positions from the UV maps will allow to associate to these galaxies reliable UV photometric redshifts. Note that the combination with UV photometric data will also help in removing from (sub-)mm and radio observations any contamination from low-luminosity, unobscured AGNs and low-$z$ star-forming galaxies.

\begin{figure}
\epsscale{1}\plotone{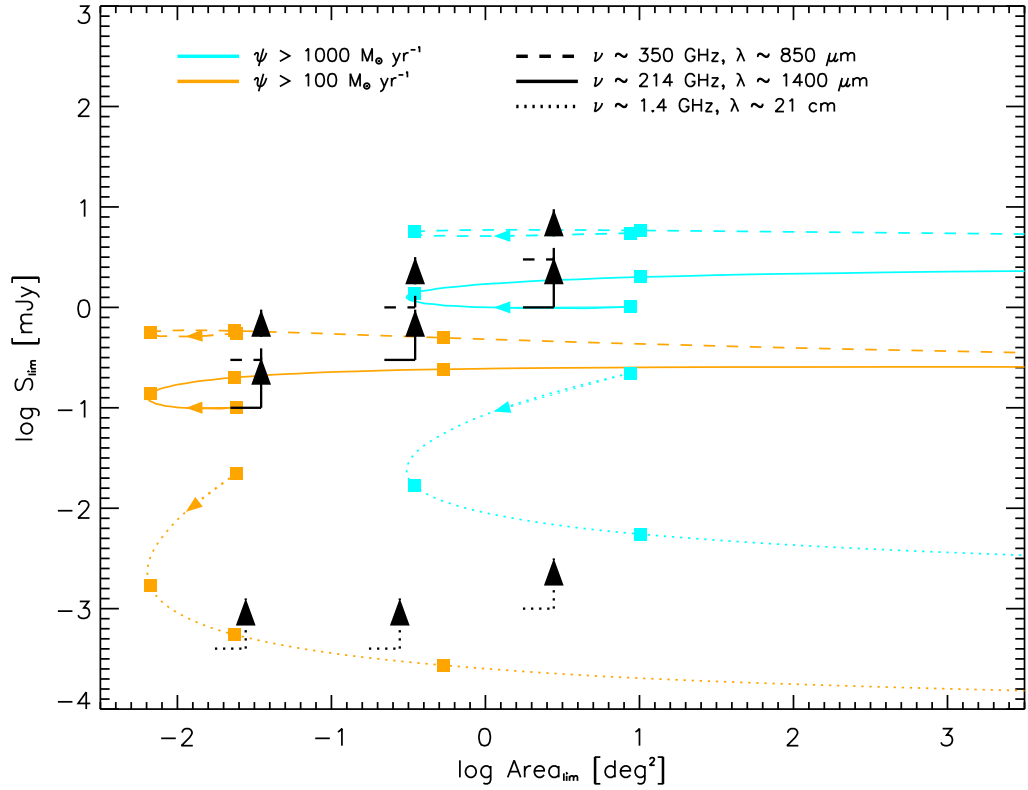}\caption{Diagram showing the limiting
flux and the area of a (sub-)mm and radio survey required to detect at least $30$ dusty galaxies per redshift bin $\Delta z\approx 1$. Results are shown for two different thresholds in intrinsic SFR $\psi\ga 100$ (orange lines) and $\ga 1000\, M_\odot$ yr$^{-1}$ (cyan lines), at three wavelengths: $850$ (dashed) and $1400\, \mu$m (solid) of interest for \textsl{ALMA}, and $21$ cm ($1.4$ GHz, dotted) of interest for \textsl{SKA}. Along each curve redshift increases following the small colored arrows, with the dots referring to $z\approx 1$, $3$ ,$5$, and $7$. The black upward arrows illustrate the \textsl{ALMA} and \textsl{SKA} $5\,\sigma$ sensitivity limits ($500$ hours of integration time, see text for details) for surveys on $100$, $1000$, and $10000$ arcmin$^2$.}\label{fig|surveydesign}
\end{figure}

Specifically, given the areas covered by current or upcoming UV surveys (Bouwens et al. 2015), in Fig.~\ref{fig|surveydesign} we present the required sensitivity  to detect at least $30$ objects (to get sound statistics) with a given SFR threshold in a redshift bin of width $\Delta z\approx 1$. In particular, we focus on SFR $\psi\ga 100$ and $1000\, M_\odot$ yr$^{-1}$, and consider three wavelengths: $850$, $1400\, \mu$m of interest for \textsl{ALMA}, and $21$ cm ($1.4$ GHz) of interest for \textsl{SKA}. The dots refer to redshift bins centered around $z\sim 1$, $3$, $5$, and $7$, with the redshift increasing following the small colored arrows. The upward black arrows illustrate the $5\sigma$ sensitivity limits of \textsl{ALMA} and \textsl{SKA}, for a total integration time of $500$ hours, on survey areas of $100$, $1000$ and $10000$ arcmin$^2$ (from left to right). Here we have adopted as reference the following specifications: for \textsl{ALMA} (in survey-mode configuration, see \texttt{http://www.ioa.s.u-tokyo.ac.jp/ $\sim$ytamura/Wiki/?plugin=attach\&refer=ALMA \& openfile=tamura-almawg-060302.pdf}) at $850-1400\, \mu$m, a field-of-view (FOV) of $0.02-0.04$ arcmin$^2$ and a $5\sigma$ sensitivity of $0.1-0.05$ mJy hr$^{-1/2}$; for \textsl{SKA} (configuration \textsl{SKA1-MID}; see Prandoni \& Seymour 2015) at $21$ cm ($1.4$ GHz) a FOV of about $0.35$ deg$^2$ and a $5\sigma$ sensitivity of $0.01$ mJy hr$^{-1/2}$.

It is seen that on $100$ arcmin$^2$, easily covered by subfields of current UV surveys, both \textsl{ALMA} and \textsl{SKA} can detect tens of galaxies at $z\approx 4-6$ with SFRs $\psi\ga 100\, M_\odot$ yr$^{-1}$; on the other hand, on $1000$ arcmin$^2$, the widest area currently surveyed in the UV, is needed to statistically sample galaxies with SFRs $\psi\ga 300\, M_\odot$ yr$^{-1}$; finally, an area of $10000$ arcmin$^2$, as possibly surveyed by the future \textsl{LSST} (e.g., Ivezic et al. 2008), will enable to detect an appreciable number of dusty galaxies with SFRs $\psi\ga 1000\, M_\odot$ yr$^{-1}$. Notice that, at the flux limits and on the survey areas considered here, gravitational lensing is only marginally effective, even at the wavelengths $\lambda\ga 1$ mm, where its effect on the counts is most relevant. For example, from Fig.~\ref{fig|diffcounts} (bottom right panel) it is seen that at $1.4$ mm for fluxes of $1$ mJy the lensed sources are only a small fraction around a few percent of the total population.

On the (sub-)mm side, another interesting instrument that could be exploited for these observations is \textsl{NIKA2}; we adopt as reference specifications (see \texttt{http://ipag.osug.fr/ nika2/Instrument.html}) at $1.2$ mm a FOV of about $40$ arcmin$^2$ and a 5$\sigma$ sensitivity around $1.3$ mJy hr$^{-1/2}$.
Then in $500$ hours it can attain a $5\sigma$ sensitivity of $0.1$ mJy on $100$ arcmin$^2$, of $0.3$ mJy on $1000$ arcmin$^2$, and of $1$ mJy on $10000$ arcmin$^2$; thus \textsl{NIKA2} at $1.2$ mm will perform similarly to \textsl{ALMA} at $1.4$ mm. On the radio side, the \textsl{SKA} precursor \textsl{MeerKAT} is also interesting; we adopt as reference specifications (see Prandoni \& Seymour 2015) at $21$ cm ($1.4$ GHz) a FOV of about $0.8$ deg$^2$ and a 5$\sigma$ sensitivity around $0.01$ mJy hr$^{-1/2}$. In $500$ hours it can attain a $5\sigma$ sensitivity of $0.5\,\mu$Jy on $100$ arcmin$^2$, of $0.5\,\mu$Jy on $1000$ arcmin$^2$, and of $0.8\,\mu$Jy on $10000$ arcmin$^2$; thus it will perform similarly to \textsl{SKA}. However, we note that \textsl{NIKA2} and \textsl{MeerKAT}, differently from \textsl{ALMA} and \textsl{SKA}, in the surveys considered above would work close to their confusion limit. Such instruments feature a resolution around $10$ and $5$ arcsec, respectively; these imply, on considering the shape of the counts (cf. Fig.~\ref{fig|diffcounts}), a confusion limit around $0.1$ mJy for \textsl{NIKA2} at $1.2$ mm, and at the $\mu$Jy level for \textsl{MeerKAT} at $21$ cm ($1.4$ GHz).

In summary, surveying common areas in the UV as well as in the (sub-)mm and/or radio bands looks a most promising strategy to characterize the SFR function at the bright end $\psi\ga 100\, M_\odot$ yr$^{-1}$ up to $z\la 8$.

\section{Summary and conclusions}\label{sec|summary}

The history of star formation in massive galaxies (the host of high-redshift
quasars) is a fundamental problem in galaxy evolution. In the present paper we address two important issues: is star formation in galaxies mainly regulated by \textit{in-situ} processes or by merging ? how does the presence of dust affect the statistics of the star formation rate in galaxies at high redshift $z\ga 3$ ?

To cast light on these issues, we have designed a method (see
Sect.~\ref{sec|SFRfunc}) to build up the intrinsic SFR function at different
redshifts up to $z\la 10$. In detail, at $z\la 3$ we have fitted a Schechter
function to the UV data for SFRs $\psi\la 30\, M_\odot$ yr$^{-1}$ and to the
far-IR data for SFRs $\psi\ga 100\, M_\odot$ yr$^{-1}$. We have further
imposed that at $z\ga 8$ the UV-inferred SFR function is representative of
the intrinsic one, since we expect small attenuation by dust due to the short age of the Universe. This allows us to set the redshift evolution of the
Schechter parameters, and hence to work out specific predictions for the SFR
functions over the full range $z\sim 0-10$.

We have found that for $z\la 7$ the UV-inferred SFR function, even when
corrected for dust-absorption according to the standard prescriptions based
on the UV slope, strongly underestimate the intrinsic SFR function for SFRs
$\psi\ga 30\, M_\odot$ yr$^{-1}$. Thus our result on the SFR function implies the existence of a galaxy population at $z\ga 4$ featuring large star
formation rates SFR $\psi\ga 10^2\, M_\odot$ yr$^{-1}$ in heavily
dust-obscured conditions. These galaxies constitute the
high-redshift counterparts of the dusty star-forming population already
surveyed for $z\la 3$ in the far-IR band by the \textsl{Herschel} space
observatory. A number of these objects have been discovered thanks to
spectroscopic follow-up of high-redshift candidates identified in UV or IR
surveys, and their corresponding number densities are well reproduced by our
intrinsic SFR function. We have further validated the latter by comparison with the observed (sub-)mm counts, redshift distributions, and cosmic infrared background (see Sect.~\ref{sec|counts}), finding an excellent agreement.

We have exploited the continuity equation approach and the `main sequence'
star-formation timescales to show that our intrinsic SFR function is fully
consistent with the stellar mass function of active, star-forming galaxies observed at redshift $z\ga 4$ (see Sect.~\ref{sec|cont}). In particular, we reproduce the considerable abundance of galaxies with stellar masses in excess of a few $10^{10}\, M_{\odot}$ at redshift $z\ga 4$, and even their still substantial number densities out to $z\sim 6$. On the contrary, we show that the UV-inferred SFR function would produce a strong deficit of galaxies with such large stellar masses.

We have computed average relationships between intrinsic SFR and stellar mass vs. halo mass via the abundance matching technique (see
Sect.~\ref{sec|abmatch}). We find that such relationships show little if no
evolution with redshift at given $M_{\rm H}$; this clearly indicates that the star formation in galaxies at high redshift $z\ga 4$ is regulated by similar, \emph{in-situ} processes, and not by merging or gas infall from cosmological
scales. We have pointed out that our results on the intrinsic SFR functions
straightforwardly overcome the `impossibly early galaxy problem' recently
pointed out by Steinhardt et al. (2015).

In order to probe the bright end of the SFR functions at $z\ga 4$, we have
computed the expected galaxy number counts and redshift distributions
(including galaxy-scale gravitational lensing) of dusty starforming galaxies. We have also designed an observational strategy (see Sect.~\ref{sec|hunting}) to hunt these galaxies based on a preselection in the far-IR or
(sub-)mm band with \textsl{Herschel} and \textsl{SCUBA-2},
possibly supplemented by on source observations with mm instruments like \textsl{AzTEC} and \textsl{ALMA}, aimed at recovering photometric (or even spectroscopic) redshifts (see Sect.~\ref{sec|FIRnature}).

We have investigated (see Sect.~\ref{sec|UVnature}) the nature of the
UV-selected galaxies at $z\ga 4$, finding that their attenuation properties
are strongly in excess with respect to those routinely estimated from the UV
slope, i.e., via the $\beta_{\rm UV}-$IRX correlation. This is because star
formation preferentially occurs within molecular clouds, i.e., cocooned
environments extremely rich in dust; on the other hand, the UV slope mostly
refers to the milder attenuation of the emission from relatively older stars
by the diffuse cirrus dust component. We have shown that a simple, powerlaw
representation of the UV attenuation due to molecular clouds in terms of the
SFR, maps the intrinsic SFR function onto the observed UV luminosity function.

We have shown that dusty, strongly star-forming galaxies with $\psi\ga 30\, M_\odot$ yr$^{-1}$ are not lost in the UV, but rather are moved by their strong attenuation $A_{\rm UV}\ga 2.3$ at fainter magnitudes, where they are outnumbered by the intrinsically faint and poorly attenuated galaxies. Such a highly star-forming, dust-obscured and massive galaxies are expected to be located on the high side of the $M_{\rm UV}-M_\star$
relationship; as such these constitute particularly suitable targets for
far-IR and (sub-)mm observations with current instruments, and for near/mid-IR observations with the \textsl{JWST}.

We have also discussed (see Sect.~\ref{sec|UVnature}) how the intrinsic SFR function at high-redshift could be probed by combining current UV surveys with observations from (sub-)mm instruments like \textsl{ALMA} and \textsl{NIKA2}, and upcoming radio facilities like \textsl{SKA} and its precursors.

As a concluding remark, we stress that collecting large statistics of UV and far-IR selected galaxies at high redshift is extremely informative on timescales for dust production and destruction. For instance, in the case of AZTEC-3 at $z\approx 5.3$ (Riechers et al. 2014) and HLFS3 at $z\approx 6.3$ (Cooray et al. 2014) far-IR data indicate SFRs $\psi\sim 1000\,M_{\odot}$ yr$^{-1}$, stellar masses $M_{\star}\sim 1-5\times 10^{10}\, M_{\odot}$, and dust masses $M_d \sim 3\times 10^8\, M_{\odot}$. The star-formation timescale $\tau_\star=M_{\star}/\psi\sim 1-5 \times 10^7$ yr implies that a large amount of dust has been rapidly accumulated in these galaxies; adopting a Chabrier IMF and no dust destruction, for type-II SN explosion a dust mass yield of $m_d\sim 0.7-3\,  M_{\odot}$ per SN is required. This yield is somewhat higher than the value found for SN 1987A $m_d\sim 0.8\, M_{\odot}$ per SN, possibly an upper bound (Matsuura et al. 2015).

On the other hand, the SN-driven shock waves destruct dust grains on a timescale $\tau_D=\tau_{\rm SN}\, M_{\rm ISM}/m_g$, where $\tau_{\rm SN}$ is the time between SN explosions, $M_{\rm ISM}$ is the mass of the ISM (gas and dust), and $m_g$ is the mass of ISM cleared per SN explosion (e.g., Slavin et al. 2014). For the Milky Way, $\tau_{\rm SN}\sim 125$ yr, $M_{\rm ISM}\sim 5\times 10^9\, M_{\odot}$ and $m_g\sim 600\, M_\odot$ hold, to yield $\tau_D\sim$ Gyr. Contrariwise, for the high-$z$ galaxies mentioned above the extremely large SFRs $\psi\sim 1000\, M_{\odot}$ yr$^{-1}$ imply $\tau_{\rm SN}\sim 0.1$ yr, making $\tau_D \la 10^7$ yr so short with respect to $\tau_\star$ as to exclude that destruction can be neglected. Additional stellar sources of dust may be at work, such as W-R stars, AGB stars and type-I SNe; the first are anyhow minor dust producers, while the second and the third can form significant dust amounts but over long timescales $\ga$ a few $\times 10^8$. Note that even accretion in molecular clouds can have an important role in dust formation (for a review, see Dwek \& Cherchneff 2010).

These instances enlighten that the issues of dust formation in high-$z$ galaxies and of dust production by type-II SN and AGB stars are still open problems (see, e.g., the discussion in Dwek \& Cherchneff 2011; Dwek et al. 2015; Mancini et al. 2015; Wesson et al. 2015). Large statistical samples of dusty starforming galaxies at $z\ga 4$ will constitute key datasets for understanding the role of the physical processes involved in dust formation and destruction.

\begin{acknowledgements}
We acknowledge the referee for useful suggestions and comments. We thank Z.-Y. Cai, G. De Zotti, G. Rodighiero, and F. Shankar for helpful
discussions, and M. Negrello and S. Eales for having shared with us their data on the counts of color-selected, candidate lensed galaxies from the \textsl{Herschel}/ATLAS survey before publication. Work financially supported from PRIN INAF 2012 `Looking into the dust-obscured phase of galaxy formation through cosmic zoom lenses in the \textsl{Herschel} Astrophysical Large Area Survey' and from PRIN INAF 2014 `Probing the AGN/galaxy co-evolution through ultra-deep and ultra-high-resolution radio surveys'. J.G.N. acknowledges financial support from the Spanish MINECO for a `Ramon y Cajal' fellowship.
\end{acknowledgements}

\clearpage

\clearpage

\clearpage

\clearpage

\clearpage

\clearpage

\clearpage

\clearpage

\clearpage

\clearpage

\clearpage

\clearpage

\clearpage

\clearpage
\begin{turnpage}
\begin{deluxetable}{lcccccccccccccccccccccccccccccccccc}
\tabletypesize{\scriptsize}\tablewidth{0pt}\tablecaption{SFR Function
Parameters}\tablehead{\colhead{Parameter} & &\multicolumn{7}{c}{Intrinsic} &
& &
\multicolumn{7}{c}{UV (dust-corrected)}\\
\\
\cline{1-1} \cline{3-9} \cline{12-18}\\
\colhead{} & &\colhead{$p_0$} & & \colhead{$p_1$} & & \colhead{$p_2$} & &
\colhead{$p_3$} & & & \colhead{$p_0$} & & \colhead{$p_1$} & & \colhead{$p_2$}
& & \colhead{$p_3$}} \startdata
$\log \mathcal{N}(z)$ & &$-2.48\pm 0.06$& &$-6.55\pm 1.17$& &$12.98\pm 3.49$& &$-8.19\pm 2.48$& & &$-1.96\pm 0.07$& &$-1.60\pm 1.44$& &$4.22\pm 3.66$& &$-5.23\pm 2.48$\\
$\log\psi_c(z)$& &$1.25\pm 0.05$& &$5.14\pm 0.60$& &$-3.22\pm 1.64$& &$-1.81\pm 1.16$ & & &$0.01\pm 0.05$& &$2.85\pm 0.94$& &$0.43\pm 2.40$& &$-1.70\pm 1.61$\\
$\alpha(z)$& &$1.27\pm 0.01$& &$2.89\pm 0.23$& &$-6.34\pm 0.66$& &$4.33\pm 0.46$& & &$1.11\pm 0.02$& &$2.85\pm 0.48$& &$-6.18\pm 1.26$& &$4.44\pm 0.83$\\
\enddata
\tablecomments{Quoted uncertainties are at $1-\sigma$ level. Fits hold in the
range SFR $\psi\sim 10^{-2}-10^4\, M_\odot$ yr$^{-1}$ and redshifts $z\sim
0-8$.}
\end{deluxetable}
\end{turnpage}

\clearpage
\begin{turnpage}
\begin{deluxetable}{ccccccccccccccccccccccccccccccccccc}
\tabletypesize{\scriptsize}\tablewidth{0pt}\tablecaption{Number of
dusty, UV-selected galaxies [in arcmin$^{-2}$]}\tablehead{\colhead{$M_{\rm UV}$} & & & \multicolumn{5}{c}{$\psi\ga 100\, M_\odot$ yr$^{-1}$} & &
& \multicolumn{5}{c}{$\psi\ga 300\, M_\odot$ yr$^{-1}$} &
& &\multicolumn{5}{c}{$\psi\ga 1000\, M_\odot$ yr$^{-1}$}\\
\cline{1-1} \cline{4-8} \cline{11-15} \cline{18-22}\\
\colhead{observed} & & &\colhead{$z=3$} & &\colhead{$z=4$} & &\colhead{$z=5$} & &\colhead{$z=6$} & & &\colhead{$z=3$} & & \colhead{$z=4$} & &\colhead{$z=5$} & &\colhead{$z=6$} & & &\colhead{$z=3$} & & \colhead{$z=4$} & & \colhead{$z=5$} & &\colhead{$z=6$}}\startdata
 -17  & & &  0.14 & & 0.09  & &  0.04  & &  0.01  & & &  0.03 & & 0.02  & &  0.006 & &  0.0006 & & &  0.002 & & 0.0008 & &  0.00008 & &  0.000001\\
 -18  & & &  0.28 & & 0.18  & &  0.08  & &  0.02  & & &  0.06 & & 0.03  & &  0.009 & &  0.001  & & &  0.005 & & 0.002  & &  0.0002  & &  0.000003\\
 -19  & & &  0.28 & & 0.18  & &  0.08  & &  0.02  & & &  0.06 & & 0.03  & &  0.009 & &  0.001  & & &  0.005 & & 0.002  & &  0.0002  & &  0.000003\\
 -20  & & &  0.24 & & 0.16  & &  0.07  & &  0.02  & & &  0.05 & & 0.03  & &  0.008 & &  0.001  & & &  0.004 & & 0.001  & &  0.0001  & &  0.000002\\
 -21  & & &  0.12 & & 0.08  & &  0.03  & &  0.01  & & &  0.03 & & 0.01  & &  0.004 & &  0.0006 & & &  0.002 & & 0.0006 & &  0.00006 & &  0.000001\\
 -22  & & &  0.03 & & 0.02  & &  0.007 & &  0.002 & & &  0.006 & & 0.003 & &  0.001 & &  0.0001 & & &  0.0003 & & 0.0001 & &  0.00001 & &  0.0000002\\
\enddata
\tablecomments{For more details, see Sect.~\ref{sec|UVnature} and Fig.~\ref{fig|UVnature}.}
\end{deluxetable}
\end{turnpage}

\end{document}